\documentclass[showkeys,aps,preprint]{revtex4}
\usepackage{amsmath}
\usepackage{txfonts} 
\usepackage{graphicx}
\usepackage[T1]{fontenc}
\usepackage[utf8]{inputenc}
\usepackage{hyperref}
\usepackage{enumitem} 
\begin{document} 
  \title{Relativistic Hydrodynamics a brief review of classical and \\quantum fluids in relativistic astrophysics}
	\author{R. F. \surname{Santos}}     
	\affiliation{Escola Municipal Jardim Glaucia, Belford Roxo, Rua P\'{e}ricles N. 15, 26195-140, Belford Roxo, RJ, Brazil	}
	\email{santosst1@gmail.com}
	\author{A. C. Amaro \surname{Faria} Jr} 
	\affiliation{Universidade Federal Tecnol\'{o}gica do Paran\'{a}, 85053-525, Guarapuava, PR, Brazil	}
	\email{atoni.carlos@gmail.com}								
	\author{L. G. \surname{Almeida} }
	\affiliation{Universidade Federal do Acre, Rodovia BR 364, Km 04, 69920-900, Rio Branco, AC, Brazil	}
	\email{luis.almeida@ufac.br}
\begin{abstract} 
   The objective of this work is to revisit fundamental aspects of relativistic hydrodynamics, aiming at the construction of a first course in relativistic hydrodynamics and its applications to astrophysics at the level of end of undergraduate course and beginning of graduate course. We aim to introduce more basic concepts of basic hydrodynamics, going through models analogous to gravity to the theory of superfluids, applying mainly to astrophysics and the cosmology of the dark universe. We review the classical hydrodynamics, Galileo symmetry and its extension to Lorentz Symmetry applied to fluids, enabling the analogy of fluids with space-time. We study the conservation of the momentum-energy tensor and the energy conditions of Hawking-Ellis. In the next sections we investigate quantum effects, in particular linked to superfluids, and we also sketch an application to dark matter. In this study, we conclude that superfluidity is one of the possible ways to quantize gravity. 
\end{abstract}	

 \maketitle

\section{Introduction}
Relativistic Fluids have received great attention from the scientific community recently, see \cite{rezz,review,ming,rbefmeu}. The works of~\cite{unhur} and~\cite{vis} opened an unexpected perspective on the study of Lorentz Symmetry in fluids. Later, the seminal work of~\cite{1982} in superfluid hydrodynamics, associated with geometric structures, would generate a fruitful and still promising line of study for models analogous to gravity, \cite{an,mg}. Although the search for privileged reference frames in general relativity, \cite{gordon}, and exotic causal structures, \cite{tacos}, is not new, the advent of the work of~\cite{unhur} and~\cite{vis} brought a new spirit to the research, which began to open a new frontier linked to phenomena of superfluids,~\cite{1982}. The works of~\cite{volo} and~\cite{Huang} were certainly exponents in this investigation, which is supported by verifications of superfluid phenomena in neutron stars, see~\cite{neutron}, theoretical models linking superfluids to the cosmological constant by~\cite{volo2} and detection of phenomena in galactic structure by~\cite{mecitou}, which can be understood as an indication that dark matter is related to superfluids. Some authors started to use the name Bose-Einstein Gravitational Condensate to model the interior of ultracompact objects, as~\cite{gbecmeu}, cosmic inflation by~\cite{gbecmeu2} and dark matter by~\cite{gbec} associating these phenomena to a type of \emph{gravitational phonon}. The hydrodynamic treatment for gravitational phenomena is also a possibility of quantizing the gravitational field through the introduction of quantum potentials, see~\cite{mand,alterquant,novelo}.

Quantum potentials are a way of introducing quantum effects in the covariant formalism, this perspective has been quite successful especially with the works of~\cite{sch} and \cite{sch2}, which are crucial in the debate regarding quantum cosmology~\cite{sch3} introducing thermodynamic potentials in such formalism. Another line that we want to highlight is called Superfluid Vacuum Theory~--~SVT~\cite{svt,svt2,zlo}~--~which has been quite successful especially in the description of the dark sector,~\cite{svt}.

\section{Non-relativistic fluid}
\subsection{Continuity Equation}
We will now demonstrate the most important equation of hydrodynamics, the continuity equation.
The mass of a fluid can be defined as follows
\begin{equation}\label{massa}
m=\int\rho{dV}
\end{equation}
where $\rho$ is the volumetric density of mass and $V$ the volume occupied by the fluid, considering the mass of a given fluid that flows through an element of area $ds$ per unit of time, we have:
\begin{equation}\label{varimassa}
\frac{dm}{dt}=\frac{d}{dt}\left(\int\rho{dV}\right).
\end{equation}

We can define the current
\begin{equation}\label{defcorre}
\vec{J}=\rho\vec{v}
\end{equation}
where $\vec{v}=\vec{v}(x,y,z,t)$ is the velocity of the fluid. If we calculate the fluid current flow through the same area element as the integral of the dot product between the current and the normal vector to the area element
\begin{equation}\label{flux}
\Phi=\int_{\mathcal{D}s}\vec{J}\cdot{d}\vec{s}.
\end{equation}

By the theorem of Gauss, we relate the integral in the area with the volume integral of the current source, represented by the divergence, 
\begin{equation}\label{tgauss}
\int_{\mathcal{D}s}\vec{J}\cdot{d}\vec{s}=\int_{\mathcal{D}V}\vec{\nabla}\cdot\vec{J}dV,
\end{equation}
where the operator $\vec{\nabla}$ is given by 
\begin{equation}
\vec{\nabla}=e_{x}\partial_{x}+e_{y}\partial_{y}+e_{z}\partial_{z}.
\end{equation}
The triplet $(e_{x},e_{y},e_{z})$ is the set of unitary vectors in the respective coordinate directions, and the operation 
\begin{equation}
\vec{\nabla}\cdot\vec {J}=\partial_{x}J_{x}+\partial_{y}J_{y}+\partial_{z}J_{z}
\end{equation}
is a dot product, representing the sources and drains of current.
Substituting~\ref{defcorre} in~\ref{flux} we get
\begin{equation}
\Phi=\int_{\mathcal{D}s}\rho\vec{v}\cdot{d}\vec{s}.
\end{equation}
Considering now the rate of mass that passes through a given unit of area per unit of time, \ref{varimassa} and taking $\vec{v}$ as a constant, we write
\begin{equation}
\frac{d\Phi}{dt}=\int_{\mathcal{D}s}\frac{d\rho}{dt}\vec{v}\cdot{d}\vec{s}.
\end{equation}
Analyzing the right side of \ref{tgauss}, we have
\begin{equation}\label{leftside}
\int\frac{d\rho}{dt}\vec{v}\cdot\vec{ds}=\frac{d}{dt}\left[\int\vec{J}\cdot\vec{ds} \right].
\end{equation}
Calculating the scalar product 
\begin{equation}
\vec{v}\cdot{d}\vec{s}=\frac{dxds}{dt},
\end{equation} 
we find 
\begin{equation}
\frac{dxds}{dt}=\frac{dV}{dt }. 
\end{equation}

We now consider the action of the operator ${d}/{dt}$ on the right side of \ref{tgauss}, 
\begin{equation}
\frac{d}{dt}\left[\int\vec{\nabla}\cdot\vec {J}dV\right]=\int\vec{\nabla}\cdot\vec{J}\frac{dV}{dt},
\end{equation} 
which is equivalent to saying that the font $\vec{\nabla}\cdot \vec{J}$ is a time-independent function. Therefore, we can write 
\begin{equation}\label{continum}
\frac{d\rho}{dt}=\vec{\nabla}\cdot\vec{J}.
\end{equation}
This is the so-called \emph{continuity equation}, which implies that fluid is a conserved quantity, so the same amount of fluid that enters the pipe is the amount that leaves, as the sources and sinks are computed in the temporal variation of the fluid density.

\subsection{Conservation of Momentum -- The Navier Stokes equation}
We are going to calculate the pressure under a fluid element. We can consider an infinitesimal element of volume of cubic form and calculate the force on walls that have a normal in the $x$ direction 
\begin{equation}\label{press}
dF_{x}=p(x,y,z)dydz-p(x+dx,y+dy,z+dz)dydz,
\end{equation}
expanding $p(x+dx,y+dy,z+dz)$ in Taylor series and rewriting~\ref{press}, we can approximate
\begin{equation}
dF(x)=p(x,y,z)dydz-p(x,y,z)dydz-\frac{\partial{p}}{\partial{x}}dxdydz.
\end{equation}
Considering then the force on the mass element $dm$ and remembering that 
\begin{align}
&	\rho=\frac{dm}{dV}\nonumber\\
&	dV=dxdydz\nonumber, 
\end{align}
we can write
\begin{equation}
(dm)\vec{a}=d\vec{F}_{x}=-\vec{\nabla}pdV,
\end{equation}
where 
\begin{equation}
\vec{\nabla}{p}=e_{x}\partial_{x}p+e_{y}\partial_{y}p+e_{z}\partial_{z}p
\end{equation}
is the pressure gradient. We know that 
\begin{align}
& m\frac{d\vec{v}}{dt}=\vec{F} \nonumber\\
& \frac{dm}{dV}\frac{d\vec{v}}{dt}=\frac{d\vec{F}}{dV}=-\vec{\nabla}p, \nonumber
\end{align}
where 
\begin{equation}
\vec{v}=\vec{v}(t,\vec{r}),
\end{equation}
we then write:
\begin{equation}\label{volta}
\frac{d\vec{v}(t,\vec{r})}{dt}=\frac{\partial\vec{v}}{\partial{t}}dt+\frac{\partial\vec{v}(t,\vec{r})}{\partial{\vec{r}}}d\vec{r},
\end{equation}
where 
\begin{equation}
\frac{\partial\vec{v}(t,\vec{r})}{\partial{r}}=\frac{\partial\vec{r}}{\partial{t}}\cdot \frac{\partial\vec{v}}{\partial\vec{r}}\vec{v}=\frac{\partial\vec{v}(t,\vec{r})}{\partial{ t}}+(\vec{v\cdot\vec{\nabla}})\vec{v}.
\end{equation}

Returning the result above in \ref{volta}, we get
\begin{equation}\label{ber}
\frac{\partial\vec{v}}{\partial{t}}+(\vec{v}\cdot\vec{\nabla})\vec{v}=-\frac{1}{\rho} \vec{\nabla}p.
\end{equation}
The equation \ref{ber} is the \emph{Navier Stokes equation}, which corresponds to Newton's second law for the fluid mass element $dm$.

\subsection{Barotropic fluids}
A fluid is defined as \emph{barotropic} if the energy density is a function only of the pressure, or $\rho\equiv\rho(p)$, as seen in~\cite{vis3}. Therefore, the inverse function shall be $p=p(\rho)$ and we can define
\begin{equation}
\mathcal{P}=\int\frac{dp}{\rho}\Rightarrow d\mathcal{P}=\frac{dp}{\rho},
\end{equation}
thus 
\begin{equation}
d\vec{x}\cdot\vec{\nabla}\mathcal{P}=\frac{dp}{\rho}, 
\end{equation}
and we can write
\begin{equation}
\frac{dp}{\rho}=d\vec{x}\cdot\frac{\vec{\nabla}p}{\rho},
\end{equation}
so
\begin{equation}
d\vec{x}\cdot\vec{\nabla}\mathcal{P}=d\vec{x}\cdot\frac{\vec{\nabla}p}{\rho}
\end{equation}
then
\begin{equation}
d\vec{x}\cdot\left(\vec{\nabla}\mathcal{P}-\frac{\vec{\nabla}p}{\rho}\right)=0
\end{equation}
indicating then that
\begin{equation}
\vec{\nabla}\mathcal{P}=\frac{\vec{\nabla}p}{\rho}.
\end{equation}
We then write the acceleration as the sum of the gravitational potential gradient and the gradient of~$\mathcal{P}$,
\begin{equation}
\vec{a}=\vec{\nabla}G-\vec{\nabla}\mathcal{P}=\vec{\nabla}\left(G-\int\frac{dp}{\rho}\right ).
\end{equation}

Returning the equation \ref{ber}, we can see that 
\begin{equation}\vec{v}\cdot\vec{\nabla}\vec{v}=\frac{1}{2}\vec{\nabla}\left(\vec{v}\cdot\vec{v}\right)-\vec{v}\times\vec{\nabla}\times\vec{v}, 
\end{equation}
we then define the vorticity of a fluid as
\begin{equation}\label{vo}
\vec{w}=\vec{\nabla}\times\vec{v}.
\end{equation}
Replacing \ref{vo} in \ref{ber}, we have
\begin{equation}
\frac{\partial\vec{v}}{\partial{t}}-\vec{v}\times\vec{w}=\vec{\nabla}\left[G-\int\frac{dp} {\rho}-\frac{|\vec{v}|^{2}}{2}\right].
\end{equation}
If we do $\vec{w}=0$ in \ref{vo} we have an irrotational fluid. If we have, in addition to an irrotational liquid, a fluid with steady flow ${\partial\vec{v}}/{\partial{t}}=0$, then we have:
\begin{equation}
\vec{\nabla}\left[G-\int\frac{dp}{\rho}-\frac{|\vec{v}|^{2}}{2}\right]=0,
\end{equation}
so we have the \emph{Bernouli equation}
\begin{equation}
\left[G-\int\frac{dp}{\rho}-\frac{|\vec{v}|^{2}}{2}\right]=constant.
\end{equation}
Considering, therefore, a steady state
\begin{equation}
f(t)=G-\int\frac{dp}{\rho}-\frac{|\vec{v}|^2}{2}-\frac{\partial\phi}{\partial{t} },
\end{equation}
and doing $f=-g, G=-gz, p=constant$, we find
\begin{equation}
G-\frac{p}{\rho}-\frac{|\vec{v}|^2}{2}-\frac{\partial\phi}{\partial{t}}=constant.
\end{equation}

\subsection{Potential flow}

A potential flow is defined according to the following relation of fluid velocity with a scalar potential
\begin{equation}\label{potescoa}
\vec{v}=\vec{\nabla}\phi,
\end{equation}
if we extract the divergence of \ref{potescoa}
\begin{equation}
\vec{\nabla}\cdot\vec{v}=\vec{\nabla}\cdot\vec{\nabla}\phi=\nabla^{2}\phi=0,
\end{equation}
we find the \emph{Laplace equation}
\begin{equation}\label{lap}
\nabla^{2}\phi=0.
\end{equation}

This type of flow is said to be incompressible. Which implies that ${d\rho}/ {dt}=0$, by the continuity equation and considering the definition of the current \ref{defcorre}, we have that 
\begin{equation}
\vec{\nabla}\cdot\vec{J} =\vec{\nabla}\cdot\rho\vec{v},
\end{equation}
using the derivative properties, we get
\begin{equation}
\vec{\nabla}\cdot\vec{J}=\vec{\nabla}\rho\cdot\vec {v}+\rho\vec{\nabla}\cdot\vec\nabla\phi,
\end{equation}
 in this case, by the equation \ref{lap} the second term vanishes, what makes the fluid incompressible is the density gradient $ \vec{\nabla}\rho=0 $. In this case, the \emph{Euler equation} comes from \ref{ber}
\begin{equation}\label{ber1}
\frac{\partial\vec{v}}{\partial{t}}=-\frac{1}{\rho}\vec{\nabla}P.
\end{equation} 
So the pressure generates acceleration of the fluid element in the flow direction.

\section{Electrodynamics as a theory of fluid}
Electromagnetism conceived current as a fluid of charge carriers, which propagated through conductors. The very idea of fields also had at its core the concept of fluid, in the case of the electric field being a fluid, the flux was calculated by Gauss' law
\begin{equation}\label{colomb}
 \vec{\nabla}\cdot{\vec{E}}=\frac{\rho}{\epsilon_{0}}
\end{equation}
and charge 
\begin{equation}
q=\int_{V}\rho{dV}. 
\end{equation}

The vortices associated with this fluid are described by Faraday's law
\begin{equation}
 \vec{\nabla}\times{\vec{E}}=-\partial_{t}\vec{B}
\end{equation}
where the magnetic field is associated with the fluid circulation property.
Furthermore, the divergence of the magnetic field is zero
\begin{equation}
 \vec{\nabla}\cdot{\vec{B}}=0,
\end{equation} 
that shows the inexistence of magnetic sources (monopoles). Completing the symmetry between the fields,
we have the Ampere-Maxwell law
\begin{equation}\label{portaq}
 \vec{\nabla}\times\vec{B}=\mu_{0}\vec{J}+\mu_{0}\epsilon_{0}\partial_{t}\vec{E},
\end{equation}
where $\mu_{0}\epsilon_{0}\partial_{t}\vec{E}$ is the displacement current discovered by Maxwell.
Here $\vec{J}= q\vec{v}$ is the charge carrier current. Applying the divergence to \ref{portaq}, we have
\begin{equation}
\vec{\nabla}\cdot\vec{\nabla}\times\vec{B}=\mu_{0}\vec{\nabla}\cdot\vec{J}+\mu_{0}\epsilon_{0 }\partial_{t}\left(\vec{\nabla}\cdot\vec{E}\right).
\end{equation}

Substituting \ref{colomb} and considering the properties of the divergence, we get
\begin{equation}
\vec{\nabla}\cdot\vec{J}+\partial_{t}\rho=0,
\end{equation}
which is the \emph{continuity equation}.

In the absence of sources $q=0,\vec{j}=0$, we have
\begin{align}
&	\vec{\nabla}\cdot{\vec{E}}=0,\label{gauss}\\
&	\vec{\nabla}\times{\vec{E}}=-\partial_{t}\vec{B},\label{lens}\\
&	\vec{\nabla}\cdot{\vec{B}}=0,\label{gaus1}\\
&	\vec{\nabla}\times\vec{B}=\mu_{0}\epsilon_{0}\partial_{t}\vec{E}.\label{amp}
\end{align}

Therefore, the vortex of an electric field implies the temporal variation of the magnetic field,
just as the magnetic vortex implies the temporal variation of the electric field.
Applying the curl (\ref{amp}), we have:
\begin{equation}
 \nabla^{2}\vec{B}-\frac{1}{c^2}\partial^{2}_{t}\vec{B}=0,
\end{equation} 
so we have the wave equation for the magnetic field. We can likewise deduce for the electric field, by applying the curl to the equation~(\ref{lens})
\begin{equation}
 \nabla^{2}\vec{E}-\frac{1}{c^2}\partial^{2}_{t}\vec{E}=0.
\end{equation} 

Although the fields are similar to fluids, the electromagnetic wave does not propagate in any medium,
this was one of the greatest revolutions in science, for it put an end to the idea of the luminiferous ether, which the Mickelson-Morley experiment later showed did not exist.

\section{Covariance of the Wave Equation}
  We now are going to check whether the Galileo transformations keep the wave equation invariant.
  \begin{equation}\label{wave}
   \nabla^{2}\phi-\frac{1}{c^2}\partial_{t}\phi=0,
  \end{equation}
where $\phi$ is a scalar field.

Performing the Galileo transformation
\begin{equation}
 \frac{\partial}{\partial{x}}=\frac{\partial{x'}}{\partial{x}}\frac{\partial}{\partial{x'}}+\frac{\partial{y'}}{\partial{x}}\frac{\partial}{\partial{x}}+\frac{\partial{z'}}{\partial{x}}\frac{\partial} {\partial{z}}+\frac{\partial{t'}}{\partial{x}}\frac{\partial}{\partial{t}}
\end{equation}
where 
\begin{equation}
 \frac{\partial{x'}}{\partial{x}}=1;\;\frac{\partial{y'}}{\partial{x}}=\frac{\partial{z'} }{\partial{x}}=\frac{\partial{t'}}{\partial{x}}=0\Rightarrow\frac{\partial}{\partial{x}}=\frac{\partial}{\partial{x'}},
\end{equation}
similarly
\begin{equation}
 \frac{\partial}{\partial{y}}=\frac{\partial}{\partial{y'}},\frac{\partial}{\partial{z}}=\frac{\partial}{ \partial{z'}}.
\end{equation}
In consequence
\begin{equation}
 \nabla^{2}=\nabla'^{2}
\end{equation}
Temporal component:
\begin{align}
 \frac{\partial}{\partial{t}}&=\frac{\partial{t'}}{\partial{t}}\frac{\partial}{\partial{t'}}+\frac{\partial{x'}}{\partial{t}}\frac{\partial}{\partial{x'}}+\frac{\partial{y'}}{\partial{t}}\frac{\partial }{\partial{y}}+\frac{\partial{z'}}{\partial{t}}\frac{\partial}{\partial{z'}}\nonumber\\
&=\frac{\partial}{\partial{t'}}-V\frac{\partial}{\partial{x'}}=\frac{\partial}{\partial{t'}}-\vec{V}\cdot{\vec{\nabla'}}.
\end{align}
The second time derivative
\begin{equation}
 \frac{\partial^2}{\partial{t}^{2}}=\left(\frac{\partial}{\partial{t'}}-\vec{V}\cdot{\vec{\nabla'}}\right)\left(\frac{\partial}{\partial{t'}}-\vec{V}\cdot{\vec{\nabla'}}\right).
\end{equation}
The transformation to the second derivative:
\begin{equation}\label{2d}
 \frac{\partial^2}{\partial{t^2}}=\frac{\partial^{2}}{\partial{t'^2}}-2\vec{V}\cdot\nabla' \frac{\partial}{\partial{t'}}+(\vec{V}).
\end{equation}
Substituting \ref{2d} in \ref{wave}, we have
\begin{equation}
 \nabla'^{2}\phi-\frac{1}{c^2}\frac{\partial^{2}\phi}{\partial{t'^2}}+\frac{2}{c ^2}\vec{V}\cdot\nabla'\frac{\partial\phi}{\partial{t'}}-\frac{1}{c^2}\vec{V}\cdot(\vec {\nabla}'\vec{V})\cdot\nabla'\phi=0.
\end{equation}

We can see that the wave equation is not covariant against Galileo transformations, in the case of sound waves, this non-covariance is associated with the presence of a medium, so the extra term represents
motion in relation to the medium, however electromagnetic waves do not depend on a medium to propagate, as demonstrated by the Michelson-Morlei experiment.

Having studied the Galileo transformation, we will now deduce a set of transformations, which are compatible with the propagation of the electromagnetic wave. Let us imagine that in a given reference frame~$S$ the propagation of a spherical wavefront is
\begin{equation}\label{esf1}
 x^{2}+y^{2}+z^{2}-(ct)^{2}=0,
\end{equation}
and in another reference frame~$S'$, the wavefront is described by
\begin{equation}\label{esf2}
 x'^{2}+y'^{2}+z'^{2}-c^{2}t'^{2}=0. 
\end{equation}

For simplicity, we can define that this signal is sent when the origin of both references coincides with $O\equiv{O}'$, in addition the reference frame$S'$ propagates with speed $V$ with respect to the reference frame~$S$. Let us assume that the transformations between the references are linear.
soon
\begin{equation}\label{t1}
 x'=A(x-Vt)
\end{equation}
and
\begin{equation}\label{t2}
 t'=Bt+Cx.
\end{equation}

These transformations must meet the following requirements
\begin{enumerate}[label=(\roman*)]
	\item A uniform rectilinear movement with respect to $S$ must also be rectilinear and uniform with respect to $S'$;
	\item For $V=0$,~($\vec{V}\equiv{V}\hat{x}$), the transformation reduces to identity; and,
	\item When a light signal is sent from $O\equiv{O'}$ in $t=t'=0$ its wavefront must propagate with $c$ in both reference frames.
\end{enumerate}

With a simple inspection we verify that \ref{t1} and \ref{t2} obey those requirements, we see that the linearity of the transformation guarantees that uniform movements in a given reference frame are mapped in uniform movements in another reference frame. We also notice that when $V=0$, we have $x=x'$. This would imply that $C\propto{V}$ and, for simplicity, we are considering that the referential~$S'$ propagates in the $x$~direction, also implying $y=y'$, $z=z'$.

Replacing \ref{t1} in \ref{esf1}
\begin{equation}\label{esf3}
 x'^{2}+y'^{2}+z'^{2}-c^{2}t'^{2}=A^{2}\left(x-Vt\right)^{2 }+y^{2}+z^{2}-c^{2}\left(Bt+Cx\right)=0,
\end{equation}
where $y^{2}+z^{2}=(ct)^{2}-x^2$.
We rewrite \ref{esf3}, highlighting $x^{2},t$ and $t^2$,
\begin{equation}\label{esf4}
 \left(A^{2}V^{2}-c^{2}B^{2}+c^{2}\right)x^{2}-2\left(A^{2}V+ c^{2}BC\right)xt+\left(A^{2}V^{2}-c^{2}B^{2}+c^{2}\right)t^{2}=0.
\end{equation}

As $x,t$ are linearly independent quantities, we have that~\ref{esf4} only vanishes if the coefficients vanish. Then we have the following system of three equations and three variables
\begin{align}\nonumber
 A^{2}V+c^{2}BC&=0,\\ A^{2}-c^{2}C^{2}&=1,\nonumber\\ B^{2}-\frac{V^2}{c ^2}A^{2}&=1\nonumber.
\end{align}
Wich we can rewrite
\begin{align}
 A^{2}&=-\frac{c^2}{V}BC,\nonumber\\ -c^{2}\frac{C}{V}(B+VC)&=1,\nonumber\\ B^{2}+VBC&=B(B+VC)=1,\nonumber
\end{align}
and therefore that $A^{2}=B^2$. Going back in the system of equations we have
\begin{equation}
 A^{2}\left(1-\frac{V^2}{c^2}\right)=1,
\end{equation}
where
\begin{equation}
 A=\pm{B}=\pm\gamma,
\end{equation}
and
\begin{equation}
 \gamma=\frac{1}{\sqrt{1-\frac{V^2}{c^2}}}.
\end{equation}
We still have
\begin{equation}
 C=-\frac{V}{c^2}\gamma,
\end{equation}
when arrive the Lorentz transformations
\begin{align}
 x'&=\gamma\left(x-Vt\right),\label{tl1}\\
 t'&=\gamma\left(t-\frac{V}{c^2}x\right),\label{tl2}\\
 y'&=y,\\
 z'&=z.
\end{align}

The inverse transforms
\begin{align}
 x&=\gamma\left(x'-Vt'\right),\\
 t&=\gamma\left(t'+\frac{V}{c^2}x'\right),
\end{align}
in the limit where $V<<c$, we reestablish the Galileo transformations.
We can now generalize the Lorentz transformation to any direction and study the transverse effects of the Lorentz transformation,
\begin{equation}
 \hat{V}\equiv\frac{\vec{V}}{V},
\end{equation}
so in the direction parallel to the motion:
\begin{align}
 \vec{x}_{\parallel}&=(\vec{x}\cdot\hat{V})\hat{V}=\frac{(\vec{x}\cdot\vec{V})\vec {V}}{V^2}\\
 \vec{x}_{\perp}&=\vec{x}-\vec{x}_{\parallel}=\vec{x}-(\vec{x}\cdot\hat{V})\hat {V},
\end{align}
and similarly
\begin{align}
 \vec{x'}_{\parallel}&=\gamma(\vec{x}_{\parallel}-\vec{V}t)\\
 x'_{\perp}&=\vec{x}_{\perp}\\
 t'&=\gamma\left(t-\frac{V}{c^2}\vec{x}\cdot\hat{V}\right)=\gamma\left(t-\frac{\vec{V }\cdot\vec{x}}{c^2}\right).
\end{align}

We take the opportunity to demonstrate the covariance of the wave equation under the Lorentz transformations
\begin{align}
 \frac{\partial}{\partial{x}}&=\frac{\partial{x'}}{\partial{x}}\frac{\partial}{\partial{x'}}+\frac{\partial{t'}}{\partial{x}}\frac{\partial}{\partial{t'}}\\
 \frac{\partial^2}{\partial{x^2}}&=\frac{\partial^{2}{x'}}{\partial{x^2}}\frac{\partial}{\partial{x'}}+\frac{\partial{x'}}{\partial{x}}\frac{\partial^2}{\partial{x'^2}}+\frac{\partial^{2}{t'}}{\partial{x^2}}\frac{\partial}{\partial{t'}}+\frac{\partial{t'}}{\partial{x}}\frac{\partial^2}{\partial{x}\partial{t}}\label{dev2x},
\end{align}
using the transformations \ref{tl1} and \ref{tl2} 
\begin{align}
\frac{\partial{x'}}{\partial{x}}&=\gamma, \\
\frac{\partial{t}}{\partial{x'}}&=\frac{V}{c^2}\gamma\\
\frac{\partial^{2}x'}{\partial{x^2}}&=0=\frac{\partial^{2}t'}{\partial{x^2}},
\end{align} 
we rewrite \ref{dev2x}
\begin{equation}\label{eq1}
 \frac{\partial^2}{\partial{x^2}}=\gamma\frac{\partial^{2}}{\partial{x'^2}}-\frac{V\gamma}{c^2}\frac{\partial^2}{\partial{x'}\partial{t}}.
\end{equation}
Following the same reasoning for the time derivative
\begin{equation}
  \frac{\partial}{\partial{t}}=\frac{\partial{t'}}{\partial{t}}\frac{\partial}{\partial{t'}}+\frac{\partial{x'}}{\partial{t}}\frac{\partial}{\partial{x'}},
\end{equation}
the second derivative:
\begin{equation}
 \frac{\partial^{2}}{\partial{t^2}}=\frac{\partial^{2}t'}{\partial{t^2}}\frac{\partial}{\partial{t'}}+\frac{\partial{t'}}{\partial{t}}\frac{\partial^{2}}{\partial{t'^2}}+\frac{\partial^{2}x'}{\partial{t}^{2}}\frac{\partial}{\partial{x'}}+\frac{\partial{x'}}{\partial{t}}\frac{\partial^{2}}{\partial{t}\partial{x'}}.
\end{equation}

Using once again the transformations~\ref{tl1} and~\ref{tl2} we have
\begin{equation}\label{eq2}
  \frac{\partial^{2}}{\partial{t^2}}=\gamma\frac{\partial^{2}}{\partial{t'^2}}-\gamma{V}\frac {\partial^2}{\partial{t}\partial{x'}},
\end{equation}
subtracting \ref{eq1} and \ref{eq2}, we get the wave equation
\begin{equation}
  \frac{\partial^{2}}{\partial{x^2}}-\frac{1}{c^2}\frac{\partial^{2}}{\partial{t^2}}= 0,
\end{equation}
and we thus demonstrate the covariance of the D'Lambertian concerning to the Lorentz transforms.

\section{The relativistic law for the composition of velocities}
Let's now calculate the instantaneous velocity of the particle, considering $x'=x'(t)$, $y'=y'(t)$, $z'=z'(t)$, we can write the velocity $v'(t)$ in the reference $S'$.
\begin{equation}
 v'_{x}=\frac{dx'}{dt'}, v'_{y}=\frac{dy'}{dt'}, v'_{z}(t)=\frac{dz }{dt}
\end{equation}
the velocity in the $S$ reference, we have:
\begin{equation}
 v_{x}=\frac{dx}{dt},v_{y}=\frac{dy}{dt},v_{z}=\frac{dz}{dt}
\end{equation}
where $x'(t)$, $y'(t)$, $z'(t)$ are related to $x(t)$, $y(t)$, $z(t)$, making the velocity reference frame parallel to the $x$ axis, or
$\vec{x}\parallel\vec{V}$, we write
\begin{equation}
 dx'=\gamma(dx-Vdt),dt'=\gamma(dt-\frac{V}{c^2}dx),dy'=dy,dz'=dz.
\end{equation}
Thus, we deduce the velocity transformations for $x$, $y$ and $z$ components, respectively:
\begin{align}
 v'_{x}&=\frac{dx'}{dt'}=\frac{dt\left(\frac{dx}{dt}-V\right)}{dt\left(1-\frac{V}{c^2} \frac{dx}{dt}\right)}=\frac{v_{x}-V}{1-\frac{v_{x}V}{c^2}},\\
 v'_{y}&=\frac{dy'}{dt'}=\frac{dy}{\gamma{dt}\left(1-\frac{V}{c^2}\frac{dx}{dt}\right)}=\frac{\sqrt{1-\frac{V^2}{c^2}}v_{y}}{1-\frac{v_{x}V}{c^2}},\\
 v'_{z}&=\frac{dz'}{dt'}=\frac{dz}{\gamma{dt}\left(1-\frac{V}{c^2}\frac{dx}{dt}\right)}.
\end{align}

\section{Space-time interval}
In relativity the spatial intervals are replaced by space-temporal intervals, because time becomes an axis, a coordinate as well as space. Therefore to describe an event in space-time, we need coordinates. To describe a point in space, thinking in two dimensions, we can build a space-time interval, which is an invariant quantity of relativity
\begin{equation}
 \delta{s}^{2}=(s_{1}-s_{0})^{2}=(x_{1}-x_{0})^{2}-c^{2}(t_{ 1}-t_{0})^{2},
\end{equation}
being the same in any inertial frame of reference
being 
\begin{equation}
   \delta{s}^{2}<0,
\end{equation}
wich we call time-type interval. This interval is related to events which are connected by a causal relationship, the signals that link these events are slower than light. We call it time interval.
If we have
\begin{equation}
 \delta{s^2}=-c^{2}\delta\tau^{2}.
\end{equation}

We may have a light-type interval,
\begin{equation}
 \delta{s}^{2}=0,
\end{equation}
where the signal connecting the two events is a light signal.
this is the boundary of the causal region. 

Finally, if we have the interval
\begin{equation}
 \delta{s}^{2}>0,
\end{equation}
this type of interval is no longer a causal interval, so the events placed by these signals cannot be
related as cause and effect, space-like region.

\begin{figure}
\centering
\includegraphics[width=5cm]{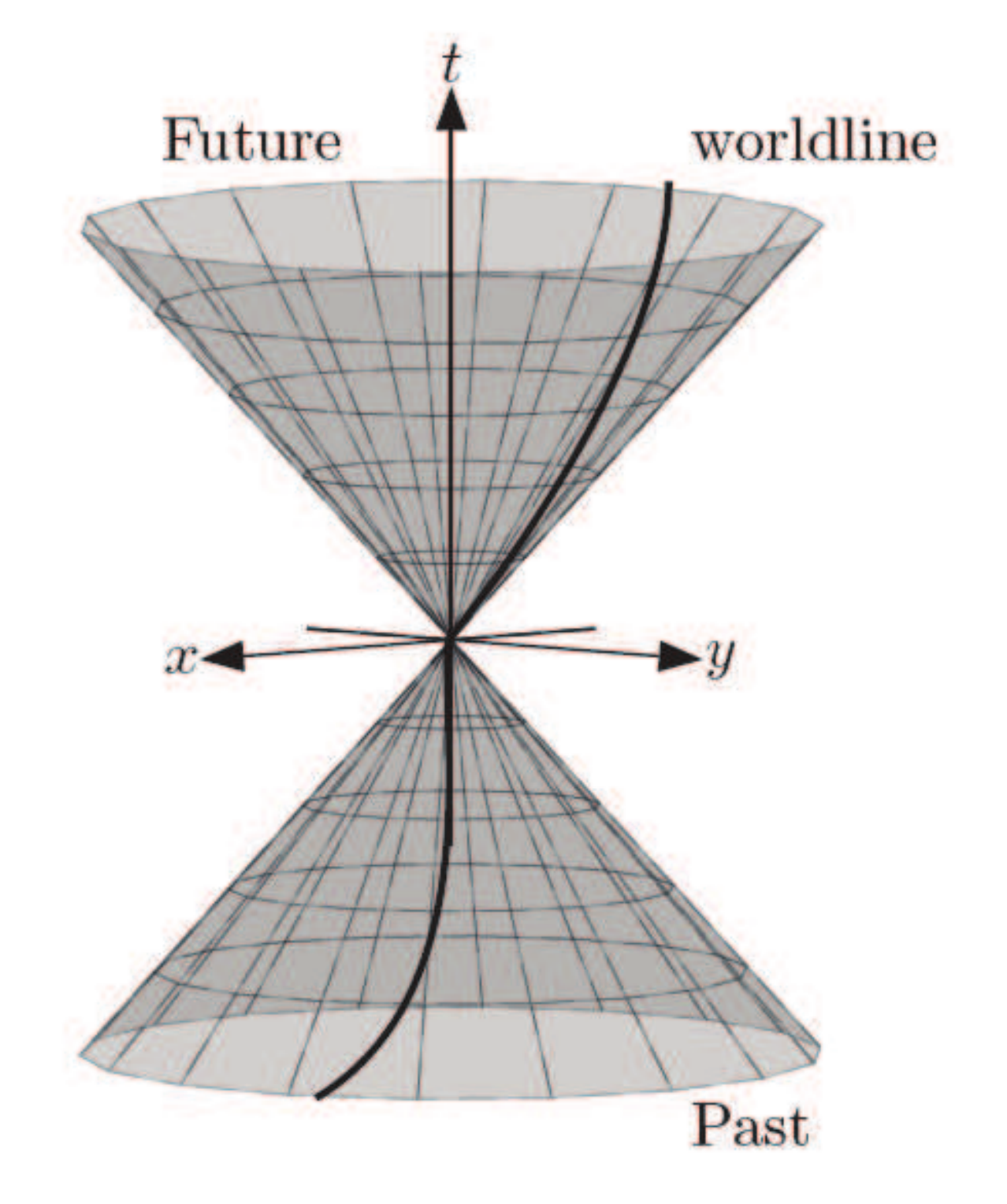}
\caption{The Cone of Light represents the causal structure of Special Relativity. The inner region represents the events, which have causal connection (time-like region). The signals are at the surface region of the cone, they propagate with the speed of light, therefore the limit of the causal zone (light-like region). In the outer zone (space-like region) the events have no causal connection, as they would be connected by supraluminal signals. }
\label{conedeluz}
\end{figure}

These properties define what we call the cone of light (see the Fig.~\ref{conedeluz}), which is the pictorial representation of the causal structure of relativity.
.

\subsection{Inertial References in Special Relativity}

  It is already known that, from the relativistic view, space and time are relativistic and interrelated quantities; therefore, the variation in one of them (eg time) in a given referential~$S'$~--~which moves with respect to another referential~$S$~--~implies the variation of the other (eg space).

  Based on the Minkowski formulation, although neither time nor space is fixed (absolute), there is a large invariant which is the ``interval
four-dimensional space-time $ds$''. This geometric formulation of relativity helps us to realize that, in fact, there is a kind of 'balancing' between categories of space and time such that $ds$ remains invariant in any inertial frame of reference, even though space and time transform separately, with the change of reference . This invariant element $ds$ is called a four-dimensional line element in Minkowki space and is written as:\\
\begin{itemize}
	\item in the reference frame~$S$
\begin{equation}\label{1}
 ds^{2}=c^{2}dt^{2}-dx^{2}-dy^{2}-dz^{2},
\end{equation}
\item and in the reference frame~$S'$ 
\begin{equation}\label{2}
 ds'^{2}=c^{2}dt'^{2}-dx'^{2}-dy'^{2}-dz'^2.
\end{equation}
\end{itemize}

  The frame $S'$ moves with velocity $v$ relative to the frame~$S$, and the time and space measurements change from~$S'$
to~$S$ and vice versa; however it comes that
\begin{equation}\label{4}
 ds^{2}=ds'^{2},
\end{equation}
because this space-temporal interval is kept invariant by referential changes.

  To better exemplify the power of synthesis of this elegant Minkowki geometric formulation, consider an observer in the frame
$S$~(in rest) and another one at the origin~$O'$ of the moving frame~$S'$, i.e.~$dx'=dy'=dz'=0 $. So, let's write the following:
\begin{equation}\label{5}
 ds'^{2}=c^{2}d\tau^{2}.
\end{equation}
In the reference frame~$S'$ the observer considers himself to be at rest, as he does not perceive his velocity in relation to the reference~$S$. While~$dt$ represents an improper time interval measured in $S$, where $dx^{2}+dy^{2}+dz^{2}=dr^{2}=v^{2}dt^2$. In fact, we have $dt\geq{d\tau}$. As~$v$ approaches~$c$, we can show that~$dt>>d\tau$, which represents an intense time dilation.
Comparing the equations~\ref{1} and~\ref{4} based on~\ref{2}, we get that:
\begin{equation}\label{6}
 c^{2}d\tau^{2}=c^{2}dt^{2}-dx^{2}-dy^{2}-dz^{2},
\end{equation}
or yet
\begin{equation}
 c^{2}dt^{2}=c^{2}d\tau^{2}+dx^{2}+dy^{2}+dz^2.
\end{equation}
Dividing by $dt^2$, we have
\begin{equation}\label{7}
 c^{2}\left(\frac{d\tau}{dt}\right)^{2}+\left(\frac{dr}{dt}\right)^{2}=c^2.
\end{equation}

The relation \ref{6} needs a closer analysis. From it we get
\begin{equation}
\left(\frac{d\tau}{dt}\right)^2=1-\frac{\left(\displaystyle{\frac{dr}{dt}}\right)^{2}}{c^2}=\left (1-\frac{v^2}{c^2}\right)
\end{equation}
then
\begin{equation}\label{8}
 dt=\frac{d\tau}{\sqrt{1-\frac{v^2}{c^2}}},
\end{equation}
which represents time dilation. So, if $v$ approaches $c$, it follows that $dt>>d\tau$, which means that proper time
(in the frame $S$), passes much slower than the improper time $dt$ measured in a frame $S$ ``stopped''.

  Through the equation~\ref{6} above, we will draw an important conclusion about the compensation (balancing), which exists between certain spatial and temporal quantities in such a way that, an invariance will be found. Let's look at the following terms
in the equation~\ref{6}:
\begin{enumerate}[label=(\roman*)]
 \item $\displaystyle{\left(\frac{dr}{dt}\right)^{2}=v^{2}}$, represents the square of the three-dimensional velocity of the particle;

 \item $c^{2}\left(\displaystyle{\frac{d\tau}{dt}}\right)^{2}=c^{2}\left(1-\displaystyle{\frac{v^2}{c^2 }}\right)=c^{2}-v^2$, this term has an abstract connotation.
Interpreted as a temporal velocity, given in the fourth dimension -- time; that is, it is a speed of time, or even a spatial analog for the degree of slowness or speed with which time flows for a moving particle. Note that
$c^{2}\left({d\tau}/{dt}\right)^{2}$ has the dimension of a squared space velocity itself.
\end{enumerate}

  Now, we can easily observe that when we add the two terms, that is, $(i)+(ii)=v^{2}+(c^{2}-v^2)=c^2$,
then we always get a constant or invariant value, which is $c^2$ by actually checking the relation \ref{6}, that is, we have
$v^{2}_{t}+v^{2}_l=c^2$, where the time velocity $v_{t}=c\sqrt{1-\frac{v^2}{ c^2}}$ and $v_{l}$ is the space velocity.
With this we can see more clearly that the sum of the squares of the space-time velocity is always constant~$c^2$
although each of them may vary from one frame of reference to another. This invariance is represented by~$c$, where
\begin{equation}\label{veloC}
	c=\sqrt{v^{2}_{t}+v^{2}_{l}}=\sqrt{c^{2}\left(\frac{d\tau}{dt}\right )^{2}+v^2}.
\end{equation}

The equation~\ref{veloC} means that all objects or particles in any frame of reference always have a space-time velocity~$c$. In this sense, we can also think as the particles in our universe are on a kind of equipotential surface~$\Phi_{u}\equiv{c^2}$ with energy~$E=m\Phi_{u}\equiv{mc^2}$, or yet using~\ref{6}, we can write 
\begin{equation}
	E=mc^{2}\left(\frac{d\tau}{dt}\right)^{2}+m\left(\frac{dr}{dt} \right)^{2}\equiv{mc^2}.
\end{equation}

\section{Fluctuations in barotropic fluids $p=p(\rho)$ under Lorentz symmetry}

Matt Visser's work \citep{vis}, detecting that Lorentz symmetry could be extended to barotropic fluids, was seminal, his conclusions gave rise to the field of study of gravity analog systems, see~\cite{unhur}, and~\cite{ vis}. Novello, Visser and Volovik even applied this analogy to black holes \cite{abh}, where they even modeled ergo-regions.
Let's now follow~\cite{unhur}, \cite{vis}, \cite{abh} to review the concept of acoustic geometry.

The speed with which an oscillation travels through a given spatial interval~$ds$ is related to the time interval spent~$dt$ as follows.
\begin{equation}\label{light}
c(\vec{x})=\frac{ds}{dt},
\end{equation}
rewriting for the space interval
\begin{equation}\label{esp}
ds=c(\vec{x})dt,
\end{equation}
where the pulse speed is given according to a property of the medium in which it propagates
\begin{equation}\label{ene}
c(\vec{x})=\frac{c_{0}}{n(\vec{x})}.
\end{equation}

The quantity $n(\vec{x})$ represents the refractive index of the medium.
Thus, the time spent traveling through the pulse~$\gamma$, is given by
\begin{equation}\label{has}
T_{\gamma}=\int\frac{ds}{c(\vec{x})}
\end{equation}
replacing \ref{light} in \ref{has} we have 
\begin{equation}
T_{\gamma}=\int\frac{ds}{c_{0}}n(\vec{x})
\end{equation}
by Fermat's principle, the states of the pulse that crosses the medium $n(\vec{x})$ are given by the minimization of the following integral , which we call the optical path
\begin{equation}\label{met}
L_{\gamma}=\int{dl}=\int{ds}n(\vec{x})
\end{equation}
starting from this minimization we can then define the trajectory of a particle in the fluid, this trajectory we will call geodesic, for that, we will define a tensor $3\times3$ called metric
\begin{equation}\label{met33}
g_{ij}=n^{2}(\vec{x})\delta_{ij}
\end{equation}
where $\delta_{ij}$ is a Kronecker delta $i,j=1,2,3$.
We have redefined \ref{met}, now we write the length differential associated with the metric
\begin{equation}\label{comp}
dl^{2}=g_{ij}(\vec{x})dx^{i}dx^{j}
\end{equation}
replacing \ref{met33} in \ref{comp} we have:
\begin{equation}
dl^{2}=n^{2}(\vec{x}(s))\delta_{ij}dx^{i}dx^{j}=n^{2}(\vec{x}(s ))ds^{2}
\end{equation}
we then rewrite \ref{met}
\begin{equation}\label{m1}
L_{\gamma}=\int{dl}=\int{ds}\sqrt{g_{ij}(\vec{x}(s))\frac{dx^{i}}{ds}\frac{dx ^{j}}{ds}}
\end{equation}
minimizing~\ref{m1}, we have the geodesic, which is the path that a free particle follows in the fluid, or even an expression of Newton's second law when the resulting forces are zero.
\begin{equation}
\frac{d}{ds}\left[\frac{g_{ij}(\vec{x}(s))\frac{dx^{i}}{ds}}
{\sqrt{g_{kl}\frac{dx^{k}}{ds}\frac{dx^{l}}{ds}}}\right]=\frac{\partial{g}_{kl} (\vec{x})}{\partial{x^{i}}}\frac{dx^{k}}{ds}\frac{dx^{l}}{ds}\left(g_{mn} (\vec{x})\frac{dx^{m}}{ds}\frac{dx^{n}}{ds}\right)^{-\frac{1}{2}}
\end{equation}
\subsection{The causal structure}
Considering the velocity measured by an observer in the laboratory \cite{abh}, we have
\begin{equation}\label{cone}
\frac{d\vec{x}}{dt}=c\vec{n}+\vec{v},
\end{equation}
taking $\vec{n}$ as a unit vector and making the squared modulus of \ref{cone}, we write
\begin{equation}
\frac{d\vec{x}^{2}}{dt^{2}}=c^{2}+2cv+v^{2},
\end{equation}
rearranging the terms
\begin{equation}
d\vec{x}\cdot{d\vec{x}}-\left(c^{2}+2cv+v^{2}\right)dt^{2}=0,
\end{equation}
or
\begin{equation}
\left(d\vec{x}-\vec{v}dt\right)^{2}-c^{2}dt^{2}=0.
\end{equation}

This is the geometry where the phonons propagate with velocity $c$, the fluid flows with velocity $\vec{v}=\vec{\nabla}\theta$, a geometry indistinguishable from the Lorentz geometry, where we write the element of line
\begin{equation}\label{metsom}
ds^{2}_{ac}=\frac{\rho}{c}\left[-c^{2}dt^{2}+||d\vec{x}-\vec{v}dt| |^2\right]
\end{equation}from where we can deduce the acoustic metric
\begin{equation}\label{I found}
\mathcal{G}^{\mu\nu}=\frac{\rho}{c}\left(\begin{array}{cc}
c^{2}-v^{2}& -\vec{v}\\
-\vec{v} & \mathcal{I}
\end{array}\right).
\end{equation}

The metric \ref{metsom} corresponds to the solution of the Klein Gordon equation
\begin{equation}\label{kg}
\Delta\psi=\frac{1}{\sqrt{-g}}\partial_{\mu}\left(\sqrt{-g}g^{\mu\nu}\partial_{\nu}\psi\right)=0.
\end{equation}

A curious property of the acoustic metric~\ref{metsom} is that the term
$||{d\vec{x}-\vec{v}dt}||$ is a module, so $d\vec{x}-\vec{v}dt>0$, implying that ${d\vec{x}}/{dt}>\vec{v}$. If we have $\vec{v}=0$ the sound propagating in a static fluid follows 
\begin{equation}
ds^{2}=\frac{\rho }{c} -c^{2}dt^{2}+dx^{2}, 
\end{equation}
then recovering the Lorentz metric, this would be known as standing waves. In another situation, if $v=c$, we have that $ds^{2}=||d\vec{x}||^{2}$, this is precisely the limit where the fluid does not allow sound propagation, the metric then coincides with the trivector modulus, being a crystallized fluid, actually a solid. On the other hand, the case in which $c<v$, we perceive the fluid flow dragging the sound waves, this would be the case that we would call \emph{ergoregion}~\cite{abh,pai2}, where rest is impossible, not there would be, in this case, standing waves.

Phonon rays propagate according to the following geodesic:
\begin{equation}\label{4tica}
\left(c^{2}-v^{2}\right)\left(\frac{dt}{ds}\right)^{2}-2v_{i}\left(\frac{dx^{i }}{ds}\right)\left(\frac{dt}{ds}\right)+1=0
\end{equation}solving the quadratic equation \ref{4tica}, we have:
\begin{equation}\label{geo}
\frac{dt}{ds}=\frac{-v_{i}\displaystyle{\frac{dx^{i}}{ds}}+\sqrt{c^{2}-v^{2}+||v_{i} \partial{dx^{i}}{ds}||^{2}}}{c^{2}-v^{2}}
\end{equation}
The equation \ref{geo} is the effective geodesic of a phonon, which propagates in the fluid. Therefore, causality properties of Lorentzian geometry can be found in material media, thus generating an interdiscipline point of contact, which can help to understand the propagation of signals in media where the gravitational field is strong, that is, it can help to better understand the gravitational collapse.

Considering the metric \ref{metsom}, we write
\begin{equation}\label{pg}
ds^{2}=\frac{\rho}{c^{2}}\left[-\left(c^{2}+v^{2}\right)dt^{2}+dx^{2 }-vdxdt\right]
\end{equation}
doing in \ref{pg} 
\begin{equation}\phi=-\frac{v^{2}(r)}{2}=\frac{2GM}{r},
\end{equation} 
which corresponds to a vortex in a fluid, and we will have
\begin{equation}\label{panle}
ds^{2}=\rho\left[\left(1-\frac{2GM}{rc^{2}}\right)dt^{2}+dx^{2}-\sqrt{\frac{2MG }{rc^{2}}}dxdt\right]
\end{equation}
The equation \ref{panle} is a Paileve-Gusland metric~\cite{pai}, a Lense-Thirring spacetime variant~\cite{pai2,vis2}.
\begin{equation}
ds^{2}_{LH}=-dt^{2}+\left[dr+\sqrt{\frac{2mG}{r}}dt\right]^{2}+r^{2}\left[ d\theta^{2}+\sin^{2}(\theta)\left(d\phi-\frac{2J}{r^{3}}dt\right)^{2}\right],
\end{equation} for the case $j=0$, being a model for the surroundings of stars and planets. The metric~\ref{panle} represents a black hole mimicker in the case of velocity $v^{2}=c^{2}{r_{h}}/{r}$, we have that~$r_{h}$ is the Schwarszchild radius where the event horizon is formed. This type of system is a good toy model for simulating quantum effects in the vicinity of a black hole. Being a contribution of hydrodynamics to the advancement of a future theory that is able to unify quantum and gravitational phenomena.

   \section{The Maxwell Tensor}

  We can define auxiliary fields, called electromagnetic potentials $\Phi$ and $\vec{A}$. Then we write the electric and magnetic fields
 in terms of potentials, \begin{equation}
 \vec{E}=-\vec{\nabla}\Phi-\frac{1}{c}\partial_{t}\vec{A}
\end{equation}\begin{equation}
 \vec{B}=\vec{\nabla}\times\vec{A}.
\end{equation} 

These potentials play an important role for example in the Aharonov-Bohm effect. We can unify the fields and build the tensor formalism. We then define the potential quadrivector
\begin{equation}
 A^{\mu}=(A^{0},A^{1},A^{2},A^{3})
\end{equation}
where $A^{0}=\Phi$, and $A^{1},A^{2},A^{3}$, are the components of the potential vector, also unifying the derivatives
$\partial_{\mu}=(c\partial_{t},\vec{\nabla})$, or $\vec{\nabla}=\partial^{i}$ where $i=1,2, 3$.

We can define the Maxwell tensor
\begin{equation}
 F^{\mu\nu}=\partial^{\mu}A^{\nu}-\partial^{\nu}A^{\mu}.
\end{equation} 

Explicitly we write
\begin{equation}
 F^{0i}=E^{i}=-\partial^{0}A^{i}-\partial^{i}A^{0}
\end{equation}
\begin{equation}
 F^{ij}=\partial^{i}A^{j}-\partial^{j}A^{i}=B_{k}\epsilon^{ijk}.
\end{equation}

We finally write the matrix associated with the Maxwell tensor 
\begin{equation}\label{tensordemax}
 F^{\mu\nu}=\left[\begin{array}{rrcccccccc}
0 & E^{1} & E^{2} & E^{3} \\
 -E^{1} & 0 & B^{3} & B^{2} \\
 -E^{2} & -B^{3} & 0 & B^{1} \\
 -E^{3} & -B^{2} & -B^{1} & 0 \\
\end{array} \right].
\end{equation}

Knowing the continuity equation \begin{equation}\label{conserve}
 \vec{\nabla}\cdot\vec{j}+c\partial_{t}\rho=0
\end{equation}
where $\vec{j}, \rho$ are the vector current density and the charge, we can generalize this current~(\ref{conserve}) using the continuity equation
\ref{conserve}\begin{equation}
 \partial_{\mu}J^{\mu}=0,
\end{equation}
then we have the quadridivergence of the quadri-current $J^{\mu}=(\rho,\vec{j})$. If we act with the derivative operator on the Maxwell tensor
\begin{equation}\label{maxwell}
 \partial_{\mu}F^{\mu\nu}=\mu_{0}J^{\nu}
\end{equation}more explicitly if we act $\partial_{i}F^{0i}=\partial_{i}E^{i}=\nabla{A}^{0}=\frac{\rho} {\epsilon_{0}}$, and yet,
$\partial_{i}F^{ij}=\mu_{0}j^{j}$ where $J^{j}=\vec{j}$.

\section{Momentum-energy tensor: relativistic fluid}

  Let's define the action following \cite{rbefmeu,Landau}
	\begin{equation}\label{action}
   S=\int\xi\left(q,\frac{\partial{q}}{\partial{x^i}}\right)dVdt=\frac{1}{c}\int\xi{d\Omega }
  \end{equation} 
	where, $q_{,i} \equiv{\partial{q}}/{\partial{x_{i}}}$ are generalized variables.

  Minimizing the action~(\ref{action})
  \begin{align}
   \delta{S}&=\frac{1}{c}\int\left(\frac{\partial\xi}{\partial{q}}\delta{q}+\frac{\partial\xi}{\partial{q_{,i}}}\delta{q_{,i}}\right)d\Omega \nonumber\\\label{ext}
 &=\frac{1}{c}\int\left[\frac{\partial\xi}{\partial{q}}\delta{q}\frac{\partial}{\partial {x^i}}\left(\frac{\partial\xi}{\partial{q_{,i}}}\right)-(\delta{q})\frac{\partial}{\partial{x ^i}}\frac{\partial\xi}{\partial{q_{,i}}}\right]d\Omega=0,
 \end{align} 
  the second term cancels out on the integration in the whole space.
 We can then write the equation of motion
 \begin{equation}\label{movi}
  \frac{\partial}{\partial{x^{i}}}\frac{\partial\xi}{\partial{q_{i}}}-\frac{\partial\xi}{\partial{q} }=0,
 \end{equation}here we assume the sum over repeated indices.
  Now following a procedure similar to the one used to verify the conservation of energy
  \begin{equation}
   \frac{\partial\xi}{\partial{x^{i}}}=\frac{\partial\xi}{\partial{q}}\frac{\partial{q}}{\partial{x^ i}}+\frac{\partial\xi}{\partial{q_{,k}}}\frac{\partial{q_{,k}}}{\partial{x^i}},
  \end{equation} substituting in the equation of motion, considering that $q_{,k,i}=q_{,i,k}$, we have:
\begin{equation}
 \frac{\partial\xi}{\partial{x^{i}}}=\frac{\partial}{\partial{x^k}}\left(\frac{\partial\xi}{\partial{ q_{,k}}}\right)q_{,i}+\frac{\partial\xi}{\partial{q_{,k}}}\frac{\partial{q_{,i}}}{\partial{x^k}}=\frac{\partial}{\partial{x^k}}\left(q_{,i}\frac{\partial\xi}{\partial{q_{,k}}} \right)
\end{equation} 
using the following property
\begin{equation}
 \frac{\partial\xi}{\partial{x^i}}=\delta^{k}_{i}\frac{\partial\xi}{\partial{x^k}}.
\end{equation}

The momentum-energy tensor in terms of the canonical variables is
\begin{equation}\label{cano}
 T^{k}_{i}=q_{,i}\frac{\partial\xi}{\partial{q_{,i}}}-\delta^{k}_{i}\xi.
\end{equation} Following this form of the momentum-energy tensor,
we can write the simplest form, by~\cite{Landau,cb},
  \begin{equation}\label{tm}
   T^{\mu\nu}=U^{\mu}P^{\nu},
  \end{equation}
where 
\begin{equation}
U^{\mu}=c\sqrt{1-\frac{v^2}{c^2}}\left[1,\frac{v^{\alpha}}{c}\right]
\end{equation}
 is the four-speed of an observer moving along with the fluid and $P^{\mu}=m_{0}cU^{\mu}$ is the quadri-momentum. We usually specify the components of the momentum-energy tensor and also define them nominally. We have the normalization
\begin{equation} \label{norm}
 g_{\mu\nu}U^{\mu}U^{\nu}=1,
\end{equation}the energy density
\begin{equation}
 T^{00}=\rho,
\end{equation}the energy flow
\begin{equation}
 T^{i0}=cP^{i},
\end{equation}
and the tension tensor
\begin{equation}
 T^{ij}=u^{i}p^{i},
\end{equation}
here Latin indices limit if $i,j=1,2,3$, Greek indices to $\mu,\nu=0,1,2,3$. The momentum-energy tensor is a symmetric matrix, that is,
\begin{equation}
 T^{\mu\nu}=T^{\nu\mu}.
\end{equation} 

Relating the momentum-energy tensor energy momentum to the kinematic structure of special relativity, we have
\begin{equation}\label{here}
 T^{\mu\nu}=m_{0}c\sqrt{1-\frac{v^2}{c^2}}U^{\mu}U^{\nu}.
\end{equation}

The conservation of the momentum tensor energy has a fundamental consequence,
\begin{equation}\label{strong1}
 \partial_{\mu}T^{\mu\nu}=m_{0}c^{2}(\partial_{\mu}U^{\mu})U^{\nu}=0.
\end{equation} 
We can define the four-force as
\begin{equation}\label{strong}
 f^{\mu}=\partial_{t}(mU^{\mu}).
\end{equation} The Lorentz force is also a four-vector
\begin{equation}\label{lore}
 f^{\nu}=\frac{q}{c}U_{\mu}F^{\mu\nu},
\end{equation} by Newton's law we arrive that
\begin{equation}\label{lore1}
 m_{0}c\partial_{t}U^{\mu}=\frac{q}{c}U_{\mu}F^{\mu\nu},
\end{equation} replacing \ref{lore1} in \ref{strong1}, we have
\begin{equation}
 \partial_{\mu}T^{\mu\nu}=\frac{q}{c}U_{\mu}F^{\mu\nu},
\end{equation} the four-current can be written as $J^{\mu}=qU^{\mu}$, and we have then
\begin{equation}
 \partial_{\mu}T^{\mu\nu}=\frac{1}{c}J_{\mu}F^{\mu\nu}.
\end{equation} 

By the equation \ref{maxwell} we arrive that 
\begin{equation}
 \partial_{\mu}T^{\mu\nu}=\frac{\mu_{0}}{c}(\partial^{\alpha}F_{\alpha\mu})F^{\mu\nu },
\end{equation} using product derivative properties
\begin{align}
 \partial^{\alpha}(F_{\alpha\mu}F^{\mu\nu})&=(\partial^{\alpha}F_{\alpha\mu})F^{\mu\nu}+ F_{\alpha\mu}\partial^{\alpha}F^{\mu\nu}\\
 \partial^{\alpha}(F_{\alpha\mu}F^{\mu\nu})&-F_{\alpha\mu}\partial^{\alpha}F^{\mu\nu}=(\partial^{\alpha}F_{\alpha\mu})F^{\mu\nu},
\end{align} 
and evoking the symmetry property
\begin{align}
(\partial^{\alpha}F_{\alpha\mu})F^{\mu\nu}&=\left[\partial^{\alpha}(F^{\mu\nu}F_{\alpha \mu})-F_{\alpha\mu}\partial^{\alpha}F^{\mu\nu}\right]\nonumber\\
  F_{\alpha\mu}\partial^{\alpha}F^{\mu\nu}&=\frac{1}{2}(F_{\alpha\mu}\partial^{\alpha}F^ {\mu\nu}+F_{\mu\alpha}\partial^{\mu}F\nu\alpha)\nonumber\\
  &=\frac{1}{2}F_{\alpha\mu}(\partial^{\alpha}F^{\alpha\mu}+\partial^{\mu}F^{\alpha\mu} )=-\frac{1}{2}F_{\alpha\mu}\partial^{\alpha}F^{\mu\alpha}\nonumber\\
	&=-\frac{1}{4}\partial^{\nu}(F_{\alpha\mu}F^{\mu\alpha}), \nonumber
	\end{align}
we have
\begin{equation}
 \frac{1}{c}F^{\mu\nu}J_{\mu}=\frac{1}{4}\left[\partial(F^{\alpha\mu}F_{\alpha\mu })-\frac{1}{4}\partial^{\nu}(F_{\mu\alpha}F^{\mu\alpha})\right],
\end{equation} then
\begin{equation}
 \partial_{\mu}T^{\mu\nu}=\frac{1}{4\pi}\partial_{\alpha}\left(F^{\nu\mu}F^{\alpha}_{ \mu}-\frac{1}{4}g^{\nu\alpha}F_{\mu\rho}F^{\mu\rho}\right).
\end{equation} 

Finally the energy-momentum tensor for the electromagnetic field is 
\begin{equation}\label{fear}
 T^{\nu\alpha}_{em}=\frac{1}{4\pi}\left(-F^{\nu\mu}F^{\alpha}_{\mu}-\frac{ 1}{4}g^{\nu\alpha}F_{\mu\rho}F^{\mu\rho}\right).
\end{equation} 
Operating with $g_{\nu\alpha}$, we easily verify that the trace of this electromagnetic tensor is null.
In a matrix, we write the energy-momentum tensor as
\begin{equation}
 T^{\mu\nu}=\left[ \begin{array}{rrcccccccc}
T^{00}& T^{01} & T^{02} & T^{03} \\
 T^{10} & T^{11} & T^{12} & T^{13} \\
 T^{20} & T^{21} & T^{22} & T^{23} \\
 T^{30} & T^{31} & T^{32} & T^{33} \\
 \end{array}\right].
\end{equation} 

Knowing the antisymmetry property of Maxwell's tensor~$F^{\mu\nu}=-F^{\nu\mu}$, easily verifiable in~\ref{tensordemax}, applying this same property in~\ref{fear} we notice that $T^{\mu\nu}=T^{\nu\mu}$. The components $T^{ij}$ with $i,j=1,2,3$ form the so-called tension tensor. The components
$T^{0i}$
\begin{equation}
 T^{0i}_{em}=-\frac{1}{4\pi}F^{0\mu}F^{i}_{\mu},
\end{equation} 
is. for the Poyting vector $\vec{S}=\vec{E}\times\vec{B}$, the directional density of energy propagation.

  \section{Isotropic fluid}

  Inspired by \ref{cano} we define the momentum-energy tensor of an ideal fluid as general as possible (without anisotropy) \cite{carrol} as
\begin{equation} 
 T^{\mu\nu}=(\rho+p)U^{\mu}U^{\nu}+pg^{\mu\nu}
\end{equation}\begin{equation}
 T^{\mu\nu}=\left[ \begin{array}{rrcccccccc}
\rho& 0 & 0 & 0 \\
 0 & -p & 0 & 0 \\
 0 & 0 & -p & 0 \\
 0 & 0 & 0 & -p \\
 \end{array}\right].
\end{equation}The momentum-energy tensor trace is
\begin{equation} \label{traco}
 T=\rho-3p,
\end{equation} and the energy density source is given by
\begin{equation}
 \rho=T^{\mu\nu}U_{\mu}U_{\nu}.
\end{equation}

We can set the projector $P_{\mu\nu}=g_{\mu\nu}-U_{\mu}U_{\nu}$,
so we find the pressure
\begin{equation}
-\frac{1}{3}P_{\mu\nu}T^{\mu\nu}=p,
\end{equation}
and the momentum-energy tensor can be rewritten in projector terms
\begin{equation}
 T^{\mu\nu}={\rho}U^{\mu}U^{\nu}+pP^{\mu\nu}.
\end{equation} 

We also have cases of a momentum-energy tensor 
\begin{equation}
T^{\mu\nu}=\rho{U^{\mu}U^{\nu}}, 
\end{equation}
which is called a dust momentum-energy, in this case we have pressure $p=0$. A radiation fluid is described by 
\begin{equation}
T^{\mu\nu}=p(U^{\mu}U^{\nu}+g^{\mu\nu} ), 
\end{equation}
and we have an equation of state $p=3\rho$. We then define the barotropic factor
\begin{equation}
 \omega=\frac{p}{\rho},
\end{equation}
where this constant factor assumes values according to the type of matter~\cite{cb}, see Table~\ref{barotropic}.
\begin{table}
\caption{Barotropic factor.}\label{barotropic}
\begin{tabular}{lr}
\hline \hline
matter & $\omega$ \\
\hline   
ordinary & $0$ \\
radiation &$ \frac{1}{3}$\\
curvature & $-\frac{1}{3}$\\
vacuum & $-1$\\
\hline
\end{tabular}
\end{table}

No imposition on the momentum-energy tensor is made \emph{a priori},~\cite{rbefmeu}.
Some fluids are repulsive, others may violate causality or are ultra-relativistic. These energy conditions classify fluids according to these criteria that we will address next.
Let a given time like vector be 
\begin{align}
	t^{\mu}&=\gamma(1,a,b,c),\\
	\gamma&=\frac{1}{1-a^{2}-b^{2} -c^{2}},
\end{align}
 with $g_{\alpha\beta}t^{\alpha}t^{\beta}$ so that $a^{2}+b^{2}+c^{2} <1$,
we have 
\begin{equation}
T_{\mu\nu}U^{\mu}U^{\nu}\geq0
\end{equation}
 and for a null vector $l^{\mu}=(1,a',b',c')$ ,$1=a'^{2}+b'^{2}+c'^{2}$, we have  
\begin{equation}
T_{\mu\nu}l^{\mu}l^{\nu}\geq0.
\end{equation}

We deduce from this that
\begin{equation}
  \rho=T^{\mu\nu}U_{\mu}U_{\nu},T_{\mu\nu}l^{\mu}l^{\nu}=(\rho+p)(U_ {\mu}l^{\mu})^2.
\end{equation}
This implies that $\rho\geq0$ and $(\rho+p)\geq0$.
Then arises the so-called \emph{Weak Energy Condition} that we will discuss next.

\subsection{Weak Energy Condition}
Let's analyze the weak energy condition
\begin{equation}
 T_{\mu\nu}t^{\mu}t^{\nu}=(\rho+a^{2}p+b^{2}p+c^{2}p)\geq0,
\end{equation}
if we do $a=b=c=0$ we have $\rho\geq0$, alternatively we do two of the null constants, for example $b=c=0$ and $a=1$
we have $(\rho+p)\geq0$. We write then
\begin{equation}\label{wec}
 \rho\geq0; (\rho+p)\geq0.
\end{equation}

We can still do  
\begin{equation}
\frac{\rho+p}{\rho}=1+\frac{p}{\rho},
\end{equation}
barotropic fluids are known from the equation of state $p=\omega\rho$,
so we rewrite the weak energy condition
\begin{equation}
 \omega\geq-1.
\end{equation}

This condition is associated with the causality of fluid flow. Therefore, the fluid flows respecting the~\cite{carrol} light cone.
There is a weaker version of this energy condition, which we will now address.
\subsection{No energy condition}

  Proceeding in the same way as for the weak energy condition \cite{carrol}, but using light-like vectors, like this
\begin{equation}
 T_{\mu\nu}l^{\mu}l^{\nu}\geq0,
\end{equation} 
where
 \begin{equation}T_{\mu\nu}l^{\mu}l^{\nu}=\rho+a'^{2}p+b'^{2}p+c'^{2}p\geq0 ,
\end{equation}
 if we do $a'^{2}+b'^{2}+c'^{2}=1$,  then
\begin{equation}\label{nec}
 \rho+p\geq0.
\end{equation}
This condition admits negative density $\rho<0$, in cases of ultra-relativistic fluids, so $l^{\mu}l_{\mu}=0$, we use light-like vectors.

\subsection{Dominant energy condition}
  An observer with four-velocity $U^{\mu}$, will see a four-velocity $-T^{\mu}_{\nu}U^{\nu}$, that is, we have that $-T^ {\mu}_{\nu}U^{\nu}$ cannot be space-like, which is equivalent to saying $T_{\mu\nu}T^{\nu}_{\lambda}t^{\mu}t^{\lambda}\leq0$ then we have
\begin{equation}
 T_{\mu\nu}t^{\mu}t^{\nu}\geq0; T_{\mu\nu}T^{\nu}_{\lambda}t^{\mu}t^{\lambda}\leq0.
\end{equation}
From $T_{\mu\nu}t^{\mu}t^{\nu}\geq0$, we get $\rho\geq0$. The current quad is not a space-like vector, which implies that $\gamma^{2}\left(-\rho^{2}+(a^{2}+b^{2}+c^{ 2})p^{2}\right)$, if we make $b=c=0$, we are left with $\rho^{2}{\geq}a^{2}p^{2}$
and $a<1$, this implies that $\rho\geq|p|$.
We finally write
\begin{equation}\label{sec}
 \rho\geq0; \rho\geq|p|.
\end{equation} 

The energy conditions above are not the only ones, we are going to state two more energy conditions, which are associated with the presence of
gravitational field. We will do this after some considerations about the momentum-energy tensor in curved spaces.

\section{The Riemman tensor and the covariant derivative}
The Riemmann tensor is the mathematical entity that measures the curvature in a Riemanian or pseudo-Riemanian manifold in the case of manifolds used in general relativity,~\cite{cb}. Let any 4-$v^{\alpha}$ be, in the absence of torsion
\begin{equation}
[\nabla_{\mu}, \nabla_{\nu}]v^{\alpha}=R^{\alpha}_{\mu\nu\beta}v^{\beta},
\end{equation}
where $[\nabla_{\mu}, \nabla_{\nu}]$ is the covariant derivative commutator. In a curved space-time it is not possible to transport a 4-vector parallel to itself, because the ``lines'' are generalized and are called geodesics, so at each point the curvature of space causes the vector to be modified, this modification then needs to be computed when we transport a 4-vector in parallel, the dot product is conserved. Therefore we can think of a dot product between a generic 4-vector $U^{\mu}$ and a base $e_{\mu}$,
\begin{equation}
    U=\mathcal{U}^{\alpha}e_{\alpha}.
\end{equation}

Studying the covariant derivative $\partial_{\nu}U=\partial_{\nu}\mathcal{U}^{\alpha}_{r}e_{\alpha}+\mathcal{U}^ {\alpha}_{r}\partial_{\nu}e_{\alpha}$, we can write the base variation $\partial_{\nu}e_{\mu}=\Gamma^{\lambda}_{\mu\nu}e_{\lambda}$. This is how we write the covariant derivative
\begin{equation}
\nabla_{\mu}U_{\nu}=\partial_{\mu}U_{\nu}-\Gamma^{\alpha}_{\mu\nu}U_{\alpha}.
\end{equation}
Now, doing
\begin{equation}\label{c1}
\nabla_{\beta}\nabla_{\mu}U_{\nu}=\nabla_{\beta}\partial_{\mu}U_{\nu}+(\nabla_{\beta}\Gamma^{\alpha} _{\mu\nu})U_{\alpha}+\Gamma^{\alpha}_{\mu\nu}\nabla_{\beta}U_{\alpha},
\end{equation}
and permuting the indices, we get
\begin{equation}\label{c2}
\nabla_{\mu}\nabla_{\beta}U_{\nu}=\nabla_{\mu}\partial_{\beta}U_{\nu}+(\nabla_{\mu}\Gamma^{\alpha} _{\beta\nu})U_{\alpha}+\Gamma^{\alpha}_{\beta\nu}\nabla_{\mu}U_{\alpha}.
\end{equation}

Subtracting \ref{c2} from \ref{c1} and considering $\Gamma^{\alpha}_{\mu\nu}=\Gamma^{\alpha}_{\nu\mu}$, which implies the absence of torsion, then we have the Riemann tensor
\begin{equation}\label{rie}
 R^{\lambda}_{\mu\nu\kappa}=\partial_{\nu}\Gamma^{\lambda}_{\mu\kappa}-\partial_{\kappa}\Gamma^{\lambda} _{\mu\nu}-\Gamma^{\lambda}_{\sigma\kappa}\Gamma^{\sigma}_{\mu\nu}+\Gamma^{\lambda}_{\sigma\nu }\Gamma^{\sigma}_{\mu\kappa}.
\end{equation}
The quantities $\Gamma^{\alpha}_{\beta\kappa}$, are known as connections, representing exactly the curvature correction when we move the vector parallel to itself. Let us then consider the metric tensor $g_{\mu\nu}$ and apply the covariant derivative to it. We therefore have
\begin{equation}\label{gama1}
\nabla_{\alpha}g_{\mu\nu}=\partial_{\alpha}g_{\mu\nu}+\Gamma^{\beta}_{\mu\nu}g_{\beta\alpha}+ \Gamma^{\beta}_{\alpha\mu}g_{\beta\nu}
\end{equation}
\begin{equation}\label{gama2}
\nabla_{\mu}g_{\nu\alpha}=\partial_{\mu}g_{\nu\alpha}+\Gamma^{\beta}_{\nu\alpha}g_{\beta\mu}+ \Gamma^{\beta}_{\mu\nu}g_{\beta\alpha}
\end{equation}
\begin{equation}\label{gama3}
\nabla_{\nu}g_{\alpha\mu}=\partial_{\nu}g_{\alpha\mu}+\Gamma^{\beta}_{\alpha\mu}g_{\beta\nu}+ \Gamma^{\beta}_{\nu\alpha}g_{\beta\mu}
\end{equation}
adding \ref{gama1} and \ref{gama2} and subtracting \ref{gama3}, remembering that in a pseudoriemanian manifold~$\nabla_{x}g_{yz}~=~0$, we get the connection written in terms of the derivatives of metric
\begin{equation}\label{conex}
\Gamma^{\beta}_{\alpha\mu}=\frac{1}{2}g^{\beta\nu}\left(\partial_{\alpha}g_{\mu\nu}+\partial_ {\mu}g_{\nu\alpha}-\partial_{\nu}g_{\alpha\mu}\right).
\end{equation}

\section{Momentum-energy tensor as source for gravitational field}
    The action of gravitational fields in general relativity is associated with the curvature of spacetime, these spacetimes are
associated with riemanine varieties \cite{cb}. The curvature of these manifolds is measured by the
Riemaann \ref{rie} that we rewrite next
\begin{equation}\label{rieb}
 R^{\lambda}_{\mu\nu\kappa}=\partial_{\nu}\Gamma^{\lambda}_{\mu\kappa}-\partial_{\kappa}\Gamma^{\lambda} _{\mu\nu}-\Gamma^{\lambda}_{\sigma\kappa}\Gamma^{\sigma}_{\mu\nu}+\Gamma^{\lambda}_{\sigma\nu }\Gamma^{\sigma}_{\mu\kappa}.
\end{equation}
We have the connection defined in the manifold under which we are working, in this case, the Christoffel symbols, which we deduce in the equation~\ref{conex}
\begin{equation}
 \Gamma^{\lambda}_{\mu\nu}=\frac{1}{2}g^{\lambda\sigma}\left(\partial_{\mu}g_{\nu\sigma}+\partial_ {\nu}g_{\sigma\mu}-\partial_{\sigma}g_{\mu\nu}\right).
\end{equation}
The Ricci tensor is found by doing the contraction $\lambda=\nu$, i.e.
\begin{equation}
 R_{\mu\kappa}=R^{\nu}_{\mu\nu\kappa}.
\end{equation}
The Ricci scalar is then
\begin{equation}
R=g^{\mu\kappa}R_{\mu\kappa},
\end{equation}
these objects characterize the curvature of space-time. We can now define the Einstein-Hilbert action \cite{rbefmeu},\cite{cb}
  \begin{equation}\label{eh}
   S_{H}=\int{R}\sqrt{-g}d^{4}x
  \end{equation}
using minimal coupling with a $S_{M}$ matter action, we have
\begin{equation}
 \delta{S}=\delta{S_{H}}+8\pi\delta{S_{M}}=0,
\end{equation}assuming $G=c=1$
ie
\begin{equation}
 \delta{S}=\int\left(\frac{\delta(\sqrt{-g}R)}{\delta{g^{\mu\nu}}}+\frac{8\pi}{\sqrt{-g}}\frac{\delta(\sqrt{-g}L_{M})}{\delta{g^{\mu\nu}}}+\frac{8\pi}{\sqrt{ -g}}\frac{\partial(\sqrt{g}L_{M})}{\partial(\partial_{k}g^{k\nu})}\delta(\partial_{k}g^{ k\mu}))\right)\delta{g}^{\mu\nu}d^{4}x=0,
\end{equation} where $\mathcal{L}_{M}=\sqrt{-g}L_{M}$ is the Lagrangian density of matter. Rearranging the terms, we have
\begin{equation}
 \delta{S}=\int\left(\frac{\delta\sqrt{-g}}{\delta{g^{\mu\nu}}}\frac{R}{2\sqrt{-g} }-\frac{\delta{R}}{\delta{g^{\mu\nu}}}+\frac{8\pi}{\sqrt{-g}}\frac{\delta(\mathcal{ L}_{M})}{\delta{g^{\mu\nu}}})\right)\sqrt{-g}\delta{g}^{\mu\nu}d^{4}x,
\end{equation}

We call the momentum-energy tensor
\begin{equation}
 T^{\mu\nu}=\frac{1}{\sqrt{-g}}\frac{\delta(\mathcal{L}_{M})}{\delta{g^{\mu\nu }}},
\end{equation}
  where 
\begin{equation}
R_{\mu\nu}=\frac{\delta{R}}{\delta{g^{\mu\nu}}}, 
\end{equation}
 we then arrive at Einstein's equation, 
\begin{equation} \label{campo0}
R^{\mu\nu}-\frac{R}{2}g^{\mu\nu}-8\pi{T^{\mu\nu}}=0.
\end{equation}

Writing the Einstein tensor 
\begin{equation} \label{campo01}
G^{\mu\nu}=R^{\mu\nu}-\frac{R}{2}g^{\mu\nu},
\end{equation}
we have
\begin{equation} \label{campo1}
G^{\mu\nu}=8\pi{T^{\mu\nu}}.
\end{equation}

By Bianchi's identity we have,~\cite{cb},
\begin{equation}
 g^{\nu\sigma}g^{\mu\lambda}\left(\nabla_{\lambda}R_{\rho\sigma\mu\nu}+\nabla_{\rho}R_{\sigma\lambda\mu\nu}+\nabla{\sigma}R_{\lambda\rho\mu\nu}\right)=0,
\end{equation}if we contract this expression, we get
\begin{equation}
 \nabla^{\mu}R_{\rho\mu}-\nabla_{\rho}R+\nabla^{\mu}R_{\rho\mu}=0,
\end{equation}
ie
\begin{equation}
 \nabla^{\mu}R_{\rho\mu}=\frac{1}{2}\nabla_{\rho}R.
\end{equation}

Then we arrive at the conservation of the momentum-energy tensor
\begin{equation} \label{campo}
 \nabla_{\mu}G^{\mu\nu}=\nabla_{\mu}\left[R^{\mu\nu}-\frac{R}{2}g^{\mu\nu}\right]=8\pi\nabla_{\mu}{T^{\mu\nu}}=0.
\end{equation}
The conservation of the momentum-energy tensor generates the so-called Tolemam-Opennhaimer-Volkov equation.

 We can now study two energy conditions, associated with gravitational regimes.
\subsection{Dominant null energy condition}
Let us now consider two energy conditions
\begin{equation} \label{ndec}
 T_{\mu\nu}l^{\mu}l^{\nu}\geq0; T_{\mu\nu}T^{\nu}_{\lambda}l^{\mu}l^{\lambda}\leq0,
\end{equation}
the first $T_{\mu\nu}l^{\mu}l^{\nu}$, has already been mentioned and implies $\rho+p\geq0$, in turn $T_{\mu\nu}T ^{\nu}_{\lambda}l^{\mu}l^{\lambda}=\rho^{2}+(a'^{2}+b'^{2}+c'^{2 })p$
. We know that $a'^{2}+b'^{2}+c'^{2}=1$, so $\rho+p\leq0$. Logically the only remaining condition is $p=-\rho$. This energy condition excludes all sources, excluded by the Dominant Energy Condition, except vacuum energy,~\cite{carrol}
\begin{equation}
 p=-\rho.
\end{equation}
From the equation \ref{tm}, we see that $T^{\mu\nu}=pg^{\mu\nu}$
The zero energy condition is associated with the cosmological constant. The solution of the Einstein equation for this type of energy-momentum tensor, implies a maximal spacetime $R_{\mu\nu}\propto{g_{\mu\nu}}$ \cite{rbefmeu},\cite{carrol},\cite{cb}.
 \subsection{Strong energy condition}
This condition is linked to attractive gravity, using the Einstein equation $R_{\mu\nu}=\left(T_{\mu\nu}-\frac{1}{2}Tg_{\mu\nu}\right )$, we have
\begin{equation}
 R_{\mu\nu}t^{\mu}t^{\nu}=\left(T_{\mu\nu}-\frac{1}{2}Tg_{\mu\nu}\right)t ^{\mu}t^{\nu}\geq0,
\end{equation} 
$T_{\mu\nu}t^{\mu}t^{\nu}\geq0$ implies $\rho+p\geq0$, 
whereas 
\begin{equation}
\frac{1}{2}Tg_{ \mu\nu}t^{\mu}t^{\nu}=-\frac{1}{2}(\rho-3p)
\end{equation}
 by the equation~\ref{traco}.

So, 
\begin{equation}
R_{\mu\nu}t^{\mu}t^{\nu}=\left(T_{\mu\nu}-\frac{1}{2}Tg_{\mu\nu}\right )t^{\mu}t^{\nu}=\rho+p+\frac{1}{2}(\rho-3p)\geq0,
\end{equation}
as $\rho+p\geq0$, we can only write that $\rho-3p\geq0$.

We finally write
\begin{equation}
 \rho+p\geq0; \rho-3p\geq0.
\end{equation} 
Violation of this condition generates repulsive gravity.
\begin{figure}
\centering
\includegraphics[width=10cm]{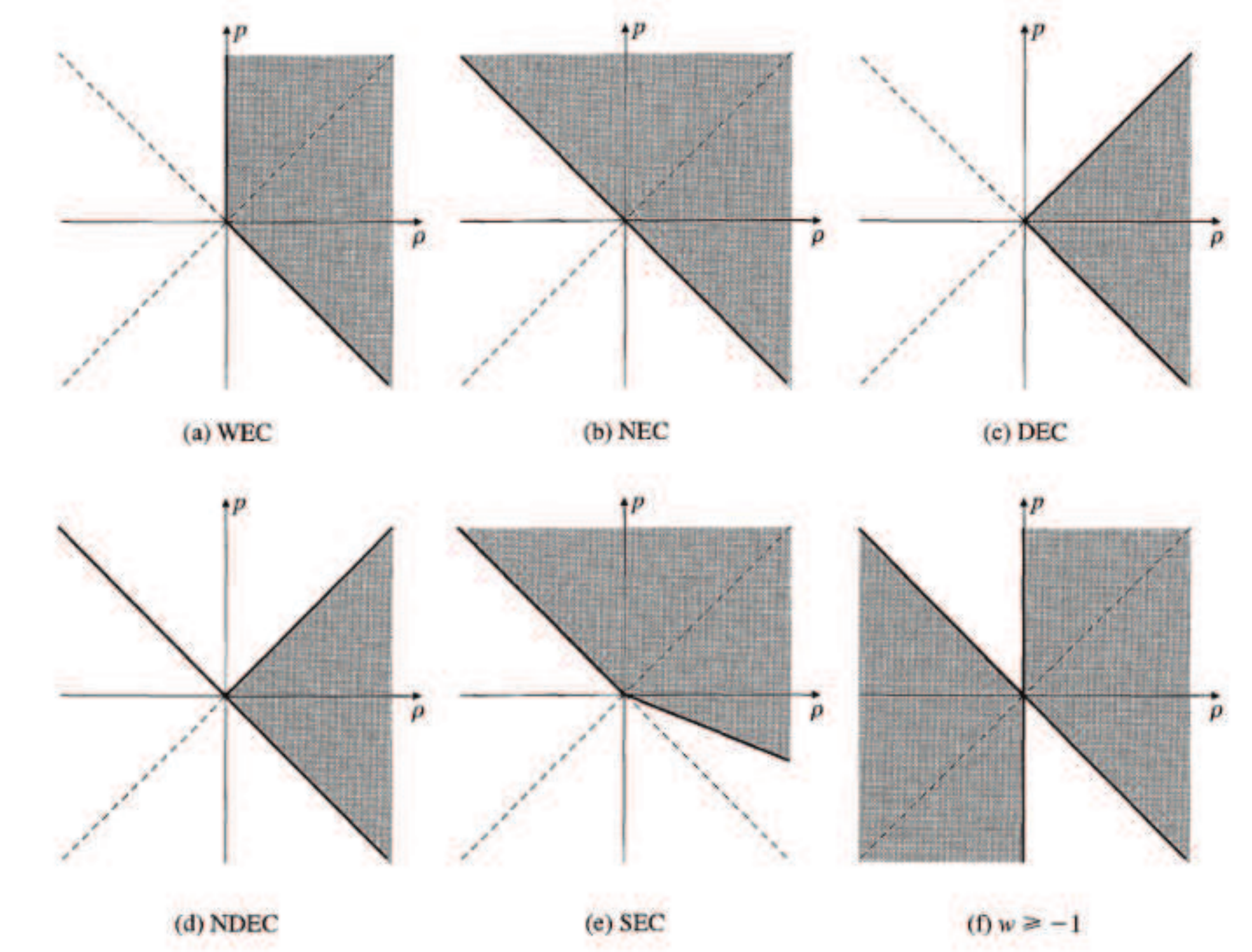}
\caption{Energy conditions are applied to perfect fluids. Representing possible values of energy density and pressure, by~\cite{carrol}.}
\label{energiac}
\end{figure}.

\subsection{Isotropic TOV demonstration: ideal fluid in a generic metric}

 We start from the equation (\ref{tm}) and derive $\nabla_{\mu}T^{\mu\nu}$, then
  \begin{equation}
   \nabla_{\mu}T^{\mu\nu}=\nabla_{\mu}\left(p+\rho\right)U^{\mu}U^{\nu}+(\rho+p)\left((\nabla_{\mu}U^{\mu})U^{\nu}+U^{\nu}\nabla_{\mu}U^{\mu}\right)+\nabla_{\mu }pg^{\mu\nu}.
  \end{equation} 
	
	The equation of continuity, or conservation of mass is 
	 \begin{equation}
   \nabla_{\mu}(\rho{U^{\mu}})=U^{\mu}\nabla_{\mu}\rho+ (\rho+p)\nabla_{\mu}U^{\mu}=0,
  \end{equation} 
then we see that
\begin{equation}
\nabla_{\mu}T^{\mu\nu}=(\nabla_{\mu}p)U^{\mu}U^{\nu}+(\rho+p)U^{\nu}\nabla_{\mu}U^{\mu}+(\nabla_{\mu}p)g^{\mu\nu}.
\end{equation}
We know that $\nabla_{\mu}U^{\nu}=\partial_{\mu}U^{\nu}+\Gamma^{\nu}_{\alpha\mu}U^ {\alpha}$, and that we have four-speed $U^{t}=\sqrt{g^{00}},U^{1}=0$.
So the only non-zero component of the 4-speed derivative is 
\begin{equation}
\nabla_{r}U^{t}=\partial_{r}(\sqrt{g^{tt}})+\Gamma^{t}_ {tr}\sqrt{g^{tt}},
\end{equation}
and we can write 
\begin{equation}
(\nabla_{\mu}p)g^{00}+(\rho+p)\sqrt{g^{00}}\nabla_{\mu}\sqrt{g^{00}}+(\nabla_{ \mu}p)g^{00}=0.
\end{equation} 
Contracting with the metric
\begin{equation}
 2\nabla_{\mu}p+(\rho+p)\sqrt{g^{00}}\Gamma^{\nu}_{00}g^{tt}g_{\mu\nu}g^{00 }=0,
\end{equation} 
we finally get
\begin{equation}
 \nabla_{\mu}p=-(\rho+p)\sqrt{g^{00}}\Gamma^{\nu}_{00}g^{00}g_{\mu\nu}(g^ {00})^{-1}.
\end{equation} 

We then arrive at the equation that represents the conservation of mass-energy
\begin{equation}
 \nabla_{\mu}p=-(\rho+p)\sqrt{g^{00}}\Gamma^{\nu}_{00}g_{\mu\nu}.
\end{equation}

\subsection{Demonstration of Isotropic TOV: metric of a spherical object}

  Let's study the Tolman-Oppenheimer-Volkoff(TOV) equation for a given spherical mass distribution. TOV is the equation that corresponds to the hydrostatic equilibrium of a relativistic fluid.
We can also interpret TOV as the conservation of the momentum-energy tensor~\cite{glen}.

The metric of a spherical distribution of mass $m$
\begin{equation}
 ds^{2}=-e^{\Phi(r)}dt^{2}+\left(1-\frac{2m}{r}\right)^{-1}dr^{2}+r ^{2}\left(d\theta^{2}+\sin(\theta)d\phi^{2}\right),
\end{equation} being 
\begin{equation}
 g^{\mu\nu}=\left[ \begin{array}{rrcccccccc}
-e^{\Phi(r)}& 0 & 0 & 0 \\
 0 & \left(1-\frac{2m}{r}\right)^{-1} & 0 & 0 \\
 0 & 0 & r^2 & 0 \\
 0 & 0 & 0 & r^{2}\sin(\theta) \\
 \end{array}\right].
\end{equation} 

We can then calculate the Christoffel symbols associated with the metric of a spherical distribution
\begin{equation}
 \Gamma^{t}_{tr}=\Phi',
\end{equation}
\begin{equation}
 \Gamma^{r}_{tt}=\Phi'e^{2\Phi}\left(1-\frac{2m}{r}\right),
\end{equation}
\begin{equation}
 \Gamma^{r}_{rr}=\frac{rm'-m}{r^{2}-2mr},
\end{equation}
\begin{equation}
 \Gamma^{\theta}_{r\theta}=\Gamma^{\phi}_{r\phi}=\frac{1}{r},
\end{equation} 
\begin{equation}
 \Gamma^{\phi}_{\theta\phi}=-\sin(\theta)\cos(\theta),
\end{equation}
\begin{equation}
 \Gamma^{r}_{\theta\theta}=\Gamma^{\theta}_{\phi\phi}=\csc^{2}(\theta),
\end{equation}
\begin{equation}
 \Gamma^{r}_{\phi\phi}=2m-r.
\end{equation} Here it must be said that we are using a system of units where mass and radius have the same unit. The Ricci tenso,r associated with these connections, has the following components
\begin{equation}
 R_{tt}=e^{2\Phi}\left[\left(\Phi''+\Phi'^{2}\right)\left(1-\frac{2m}{r}\right)+\Phi'\left(\frac{2r-3m-rm'}{r^2}\right)\right],
\end{equation}
\begin{equation}
 R_{rr}=\left(1-\frac{2m}{r}\right)^{-1}\left[\frac{(rm'-m)(2+rm')}{r^3}\right]-\Phi''-\Phi'^2,
\end{equation}
\begin{equation}
 R_{\theta\theta}=\csc^{2}(\theta)R_{\phi\phi}=(2m-r)\Phi'+m'+\frac{m}{r}.
\end{equation} 

The Ricci's scalar,
\begin{equation}
 R=g^{\mu\nu}R_{\mu\nu}=2\left[\frac{2m'}{r}+\Phi'(3m-2r+rm')-\left(1-\frac{2m}{r}\right)\left(\Phi''+\Phi'^{2}\right)\right].
\end{equation} Using the $rr$ component of the Einstein equation, 
\begin{equation}\label{einrr1}
 G_{rr}=\frac{2}{r}\left(\Phi'-\frac{m}{1-\frac{2m}{r}}\right)=\frac{8\pi{p} }{1-\frac{2m}{r}} ,
\end{equation} we then find \begin{equation}
\Phi'=\frac{m+4{\pi}r^{3}p}{r(r-2m)}.
\end{equation}
The $tt$ component is
\begin{equation}
 G_{tt}=\frac{2m'e^{2\Phi}}{r^2}=8\pi\rho{e^{2\Phi}}
\end{equation}
where $m'=\frac{dm}{dr}=4\pi\rho{r^2}$, is the continuity equation.

We can write the momentum-energy tensor \begin{equation}
 T^{\mu\nu}=\left[ \begin{array}{rrcccccccc}
\rho{e}^{-2\Phi}& 0 & 0 & 0 \\
 0 & (1-\frac{2m}{r})p & 0 & 0 \\
 0 & 0 & \frac{p}{r^2} & 0 \\
 0 & 0 & 0 &p\frac{\csc^{2}(\theta)}{r^2} \\
 \end{array}\right].
\end{equation} 

Since the functions are dependent on the $r$ coordinate, we have the covariant derivative of the momentum-energy tensor,
\begin{align}
 \nabla_{r}T^{r\nu}&=\partial_{r}T^{r\nu}+T^{\sigma\mu}\Gamma^{r}_{\sigma\mu}+T^ {r\sigma}\Gamma^{\nu}_{\sigma\nu} \\
&=\partial_{r}T^{rr}+T^{rr}(\Gamma^{\nu}_{r\nu}+\Gamma^{r}_{rr})+ T^{\theta\theta}\Gamma^{r}_{\theta\theta}+T^{\phi\phi}\Gamma^{r}_{\phi\phi}\nonumber\\
&=\left(1-\frac{2m}{r}\right)\left[\frac{dp}{dr}+(\rho+p)\Phi'\right]=0\nonumber
\end{align} 
so rises the TOV equation
\begin{equation} \label{tov}
 \frac{dp}{dr}=-(\rho+p)\frac{m+4{\pi}pr^3}{r(r-2m)},
\end{equation} 
together with the continuity equation,
\begin{equation}
 \frac{dm}{dr}=4\pi\rho{r^2},
\end{equation} 
form the so-called structure equations for a~\cite{glen} star.
 
\section{Anisotropic fluid}
 We define now the momentum-energy tensor of an anisotropic ideal gas, as seen in~\cite{rbefmeu,carrol,cb,fatima},
  \begin{equation}
   T^{\mu\nu}=(\rho+p_{t})U^{\mu}U^{\nu}+pg^{\mu\nu}+(p_{t}-p)s^ {\mu}s^{\nu},
  \end{equation} 
	where $s^{\mu}$ is perpendicular to the four-velocity fluid flow~$U^{\mu}=(1,0,0,0)$, $s^{\mu} U_{\mu}=0$.

In matrix terms
\begin{equation}
 T^{\mu\nu}=\left[ \begin{array}{rrcccccccc}
\rho& 0 & 0 & 0 \\
 0 & p & 0 & 0 \\
 0 & 0 & p_{t} & 0 \\
 0 & 0 & 0 & p_{t} \\
 \end{array}\right].
\end{equation}

The element
\begin{equation}
 \Delta=p_{t}-p,
\end{equation} which is the so-called \emph{anisotropic factor}, $\Delta>0, p_{t}>p$, that is, a repulsive factor~-- or $\Delta<0, p_{t}<p$,
in this case the anisotropy collaborates with the gravitational action.
The trace of the momentum-energy tensor is
\begin{equation}\label{traco1}
 T=\rho-p-2p_{t}.
\end{equation} 

We will now study the limitations known as energy conditions~\cite{rbefmeu,fatima}.
We will considerate energy conditions similar to the conditions for the isotropic case,
similar in the sense that time-like and light-like vectors will be chosen to suit the momentum-energy tensor
with different components.

\subsection{Low energy condition}
As in the previous sections the weak energy condition is 
\begin{equation}
 T_{\mu\nu}t^{\mu}t^{\nu}\geq0,
\end{equation}
being $t^{\mu}$ time-like vector, we establish that $t^{\mu}=(1,a,b,c)$, then $ T_{\mu\nu}t^{\mu} t^{\nu}=\rho+ap+bp_{t}+cp_{t}$. Choosing $a=b=c=1$ we arrive at the positivity of the trace
\begin{equation}
 \rho+p+2p_{t}\geq0.
\end{equation} Here the barotropic factor is modified,~$1+\omega+2\frac{p_{t}}{\rho}\geq0$
\begin{equation}
\omega\geq-1-2\frac{p_{t}}{\rho},
\end{equation} the causality is then modified, taking into account the tangential pressure.

\subsection{null energy condition}
Using the null vectors $l^{\mu}=(1,a',b',c')$ the null energy condition is
\begin{equation}
 T_{\mu\nu}l^{\mu}l^{\nu}\geq0.
\end{equation} 

We verify that, $T_{\mu\nu} l^{\mu} l^{\nu}=\rho+a'^{2}p+(b'^{2}+c'^{2})p_{t}$, choosing $b'=c'=0$, obligatorily we have $a'=1$, and we get $\rho+p\geq0$. Alternatively we make $a'=0$ and we are left with $\rho+p_{t}\geq0$, therefore
\begin{equation}
 \rho+p\geq0,\rho+p_{t}\geq0.
\end{equation} 

We obtained two similar inequalities, which represent conditions for ultra-relativistic fluids.

\subsection{Dominant energy condition}
The dominant energy condition is
\begin{equation}
 T_{\mu\nu}t^{\mu}t^{\nu}\geq0,T_{\mu\nu}T^{\nu}_{\lambda}t^{\mu}t^{\lambda}\leq0.
\end{equation} From $T_{\mu\nu}t^{\mu}t^{\nu}\geq0$, we get $\rho\geq0$. The current quad not being a space-like vector implies that $\gamma^{2}\left(-\rho^{2}+(a^{2}p^{2}+(b^{2}+ c^{2})p_{t})\right)$, if we make $b=c=0$, we are left with $\rho^{2}{\geq}a^{2}p^{2}$
and $a<1$, implies that $\rho\geq|p|$. Following a similar reasoning $a=0$, we arrive at $\rho\geq|p_{t}|$.
We finally write
\begin{equation}
 \rho\geq0; \rho\geq|p|; \rho\geq|p_{t}|.
\end{equation}
\subsection{Dominant null energy condition}
The dominant null energy condition is expressed
\begin{equation} \label{ndec1}
 T_{\mu\nu}l^{\mu}l^{\nu}\geq0; T_{\mu\nu}T^{\nu}_{\lambda}l^{\mu}l^{\nu}\leq0,
\end{equation} 
where $T_{\mu\nu}l^{\mu}l^{\nu}$, has already been calculated and implies $\rho+p\geq0,\rho+p_{t}\geq0$, in turn $T_{\mu\nu}T^{\nu}_{\lambda}l^{\mu}l^{\nu}=\rho^{2}+a'^{2}p+ (b'^{2}+c'^{2})p_{t}$. We know that $a'^{2}+b'^{2}+c'^{2}=1$, making $b'=c'=0$, so $\rho+p\leq0$. Logically the only remaining condition is $p=-\rho$. We can still make $a'=0$, implying that $\rho+p_{t}\leq0$.
So we also have $p_{t}=-\rho$, so the vacuum energy is not anisotropic. In summary
\begin{equation}
 p_{t}=p=-\rho
\end{equation} expresses the results.

\subsection{Strong energy condition}
This condition is linked to the attractive gravity, using the Einstein equation we have 
\begin{align}
R_{\mu\nu}&=\left(T_{\mu\nu}-\frac{1}{2}Tg_{\mu\nu}\right)\nonumber\\
 R_{\mu\nu}t^{\mu}t^{\nu}&=\left(T_{\mu\nu}-\frac{1}{2}Tg_{\mu\nu}\right)t ^{\mu}t^{\nu}\geq0,
\end{align} 
again $T_{\mu\nu}t^{\mu}t^{\nu}\geq0$, implies $\rho-p-2p_{t}\geq0$, already $\frac{ 1}{2}Tg_{\mu\nu}t^{\mu}t^{\nu}=-\frac{1}{2}(\rho-p-2p_{t})$ by the equation \ref{traco1}.
So $R_{\mu\nu}t^{\mu}t^{\nu}=\left(T_{\mu\nu}-\frac{1}{2}Tg_{\mu\nu}\right )t^{\mu}t^{\nu}=\rho+p+2p_{t}+\frac{1}{2}(\rho-p-2p_{t})\geq0$, as $\rho+p+2p_{t}\geq0$ we can only write that $\rho-p-2p_{t}\geq0$.
We finally write
\begin{equation}
 \rho+p+2p_{t}\geq0; \rho-p-2p_{t}\geq0.
\end{equation} The violation of this condition generates repulsive gravity.

\subsection{Demonstration of anisotropic TOV}

  As we did in the isotropic case, we write the metric here in terms of the functions $\nu(r), \lambda(r)$.
\begin{equation}
 ds^{2}=-e^{\nu(r)}dt^{2}+e^{\lambda(r)}dr^{2}+r^{2}(d\theta^{2} +\sin^{2}(\theta)d\phi^{2}),
\end{equation} 
we can now use the Einstein equation \ref{campo}. The $tt$ component is
\begin{equation}
 e^{-\lambda(r)}\left(\frac{\lambda'}{r}-\frac{1}{r^2}\right)+\frac{1}{r^2}=8 \pi\rho,
\end{equation}
and the $rr$ component
\begin{equation}\label{einrr2}
 e^{-\lambda(r)}\left(\frac{1}{r^2}+\frac{\nu'}{r}\right)-\frac{1}{r^2}=8 \pi{p}.
\end{equation}
for angular components~$\theta\theta$, it generates
\begin{equation}
 \frac{1}{2}\left(\frac{1}{2}(\nu')^{2}+\nu''-\frac{1}{2}\lambda'\nu'+\frac{1}{r}(\nu'-\lambda')\right)=8\pi{p_{t}}.
\end{equation}

The equation for $p_{t}$ represents the anisotropic pressure, here the difference with the isotropic case is manifested.
This is the usual model for anisotropic strange stars.

  Let us consider the hydrostatic equilibrium~\cite{rbefmeu,glen}
\begin{equation}
 F_{g}+F_{hydro}+F_{ani}=0,
\end{equation} where $F_{g}=\frac{1}{2}(\rho+p)d_{r}\nu(r)$ is the gravitational force, $F_{hydro}=-\frac {dp}{dr}$ the hydrostatic force
and the anisotropic force $F_{ani}=2\frac{(p_{t}-p)}{r}$.

We can then write the TOV in its anisotropic version as
\begin{equation}
 \frac{dp}{dr}=-m\frac{\rho+p}{r^2}\frac{d\nu(r)}{dr}+\frac{2}{r}(p_{t }-P).
\end{equation}
The first installment of the TOV is identical to the isotropic case, as the equations~\ref{einrr1} are very similar to the equation~\ref{einrr2},  $\nu'(r)=\Phi'(r)$. We have finally written the TOV,
\begin{equation} \label{tovani}
 \frac{dp}{dr}=-(\rho+p)\frac{m+4{\pi}pr^3}{r(r-2m)}+\frac{2}{r}(p_{ t}-p),
\end{equation}
which is the isotropic TOV.

\section{The hydrostatic stability in a De-Sitter geometry}
  It is also possible to construct a momentum-energy tensor taking into account the interaction of matter with the cosmological constant.
In this case the Einstein equation is modified to
\begin{equation} \label{const}
 R_{\mu\nu}-\frac{R}{2}g_{\mu\nu}+\Lambda{g}_{\mu\nu}=8\pi{T_{\mu\nu}},
\end{equation}
the homogeneous version of this equation would be the Einstein equation for vacuum, known as the De Sitter equation \cite{rbefmeu, carrol, cb},
\begin{equation} \label{const0}
 R_{\mu\nu}-\frac{R}{2}g_{\mu\nu}+\Lambda{g}_{\mu\nu}=0,
\end{equation}
which has as a solution
\begin{equation}
 R_{\mu\nu}=\Lambda{g}_{\mu\nu},
\end{equation}
a metric-proportional Ricci tensor. We could alternatively think about a metric-proportional energy-momentum tensor $T_{\mu\nu}=\Lambda{g}_{\mu\nu}$,
this condition is associated with the De Sitter metric
\begin{equation}
 ds^{2}=-\left(1-\frac{\Lambda{r^2}}{3}\right)dt^{2}+\frac{1}{1-\frac{\Lambda{r ^2}}{3}}dl^{2} +r^{2}(d\theta^{2}+\sin(\theta)d\phi^{2}),
\end{equation}
and the momentum-energy tensor proportional to the metric. That implies an equation of state~\ref{tm}
\begin{equation}
 p=-\rho,
\end{equation}
see \cite{carrol},\cite{cb}. Having a barometric factor $\omega=-1$, an equation of state of this type obeys the NDEC energy condition.

  To effectively calculate the TOV we need a metric~\cite{rbefmeu}.
\begin{equation}
ds^{2}=e^{\nu}dt^{2}-e^{\lambda(r)}dr^{2}-r^{2}\left(d\theta^{2}+\sin^{2}(\theta)d\phi^{2}\right),
\end{equation}
explaining the components of the metric $g_{tt}=e^{\nu},g_{rr}=-e^{\lambda}, g_{\theta\theta}=-r^{2}, g_{\phi \phi}=-r^{2}\sin^{2}(\theta)$
\begin{equation}
 g^{\mu\nu}g_{\nu\lambda}=\delta^{\mu}_{\lambda}.
\end{equation}

We then write the metric in matrix form
\begin{equation}
 g^{\mu\nu}=\left[ \begin{array}{rrcccccccc}
e^{-\nu}& 0 & 0 & 0 \\
 0 & -e^{\lambda} & 0 & 0 \\
 0 & 0 & -\frac{1}{r^2} & 0 \\
 0 & 0 & 0 &\frac{1}{(r\sin{\theta})^{2}} \\
 \end{array}\right],
\end{equation} 
here we used the metric functions $\nu=\nu(r)$ and $\lambda=\lambda(r)$. We calculate the Ricci tensor, $R_{\mu\nu}=\partial_{\alpha}\Gamma^{\alpha}_{\mu\nu}-\partial_{\nu}\Gamma^{\alpha}_{ \mu\alpha}+\Gamma^{\alpha}_{\beta\alpha}\Gamma^{\beta}_{\mu\nu}-\Gamma^{\alpha}_{\beta\mu}\Gamma^{\beta}_{\alpha\nu}$,
always remembering that $R^{\alpha}_{\beta}=g^{\alpha\mu}R_{\mu\beta}$. We can write the components of the Einstein equation:

$tt$ component,
\begin{align}
 G^{t}_{t}&=-8\pi{T^{t}}_{t},\nonumber\\
 \label{temporal}
 8\pi\rho&=e^{-\lambda}\left(\frac{\lambda'}{r}-\frac{1}{r^2}\right)+\frac{1}{r^2 }-\Lambda;
\end{align}

$rr$ component,
\begin{align}
 G^{r}_{r}&=8{\pi}T^{r}_{r},\nonumber\\
 \label{radial}
 8\pi{p}&=e^{-\lambda}\left(\frac{\nu'}{r}+\frac{1}{r^2}\right)-\frac{1}{r^ 2}+\Lambda.
\end{align} 

For a static star we consider
\begin{equation} \label{this}
 \frac{d\rho}{dt}=\frac{dp}{dt}=0.
\end{equation}
Thus, the covariant derivative of the momentum-energy tensor induces the following result
\begin{equation}
 (\rho+p)(\partial_{\mu}u_{\sigma})u^{\mu}+\partial_{\sigma}p+\partial_{\mu}pu^{\mu}u_{\sigma} =0.
\end{equation} 

We remember the $rr$ component of the metric $g_{rr}=-e^{\lambda}$, so
\begin{equation}
 (\rho+p)(\partial_{\mu}u_{t})u^{\mu}-\partial_{r}p=0,
\end{equation} like this
\begin{equation} \label{nu}
 \partial_{r}p=-\frac{(\rho+p)}{2}\frac{d\nu}{dr}.
\end{equation} We now define the mass for a spherical shell of radius $r$ as
\begin{equation}
 m(r)=4\pi\int^{r}_{0}\rho(r')r'^{2}dr',
\end{equation} differentiating the mass and substituting in the temporal component of the Einstein equation \ref{temporal} we obtain
\begin{equation}
 \frac{dm}{dr}=-\frac{1}{2}\frac{d}{dr}\left[e^{-\lambda}r-r+\frac{\Lambda{r^3}} {3}\right],
\end{equation} integrating we have
\begin{equation}
 2\int^{r}_{0}\frac{dm}{dr}dr=-\int^{r}_{0}\frac{1}{2}\frac{d}{dr}\left [e^{-\lambda}r-r+\frac{\Lambda{r^3}}{3}\right]dr,
\end{equation} so we finally arrive at the $rr$ component of the metric
\begin{equation} \label{lamb}
 e^{-\lambda(r)}=1-\frac{2m(r)}{r}-\frac{\Lambda{r^3}}{3},
\end{equation} adding the radial and temporal components of the Einstein equation, respectively the equations~\ref{temporal} and~\ref{radial}, we have,
\begin{equation} \label{soma}
 8\pi(\rho+p)=\frac{\lambda'e^{-\lambda}}{r}+\frac{e^{-\lambda}\nu'}{r}.
\end{equation}

Deriving the equation \ref{lamb} with respect to the radial coordinate we have
\begin{equation} \label{deri}
 \lambda'e^{-\lambda}=\frac{2m'}{r}-\frac{2m}{r^2}+\frac{2\Lambda{r}}{3},
\end{equation} 
we now substitute the equations \ref{deri} and~\ref{lamb} in the equation~\ref{soma}, and we get
\begin{equation} \label{ptov}
 8\pi(\rho+p)=\left(\frac{2m'}{r}-\frac{2m}{r^3}+\frac{2\Lambda}{3}\right)+\frac {1}{r}\frac{d\nu}{dr}\left(1-\frac{2m}{r}-\frac{\Lambda{r^2}}{3}\right).
\end{equation} 

Substituting \ref{nu} in the equation \ref{ptov}, we finally arrive at TOV with cosmological constant,
\begin{equation} \label{tovcos}
 \frac{dp}{dr}=-\rho\left(1+\frac{p}{\rho}\right)\frac{m+4\pi{p}r^{3}-\frac{\Lambda{r^3}}{3}}{r^{2}(1-\frac{2m}{r}-\frac{\Lambda{r^2}}{3})}.
\end{equation} 

The introduction of the cosmological constant generates an interesting effect,
if we make $m=0$, we have an effect associated with vacuum pressure
\begin{equation} \label{tovacuo}
 \frac{dp}{dr}=-\rho\left(1+\frac{p}{\rho}\right)\frac{4\pi{p}r^{3}-\frac{\Lambda{ r^3}}{3}}{r^{2}(1-\frac{\Lambda{r^2}}{3})}.
\end{equation}

We usually consider that the vacuum equation of state is~$p=-\rho$ , obeying the energy condition~\ref{ndec}, would induce the above equation to be~\ref{tovacuo} identically null, so the pressure of the vacuum was a constant.
However, if we consider the effect of the cosmological constant term as an extra pressure, similarly to what we do in the presence of tangential pressure, we could consider the part associated with the cosmological constant ($\Lambda>0$) to be an anisotropy.
An extra negative pressure, which attempts to compensate for the effects of radial pressure.

\section{Hydrodynamics with quantum potential}

\subsection{ Mandelung formalism}
The interpretation of Bohn-De's bloglie Quantum Mechanics bears a strong resemblance to a theory of fluids. Being the wave function a current density~\cite{mand} is notable the presence of a continuity equation. One way to introduce quantum effects in a relativistic scenario is through the Mandelung formalism~\cite{mand}. Where precisely we have a quantum potential introduced.

Madelung's formalism can be a good working reference for the mesoscale~\cite{mand} classical/quantum description describing the interaction between classical and quantum phenomena.

Following Chiareli~\cite{chia}, we now write Klein Gordon's equation for a function~$\psi_{m}$, which is a mass field,
 \begin{equation}\label{mande}
     \partial_{\mu}\partial^{\mu}\psi_{m}=-\frac{m^{2}c^2}{h^2}\psi_{m}.
 \end{equation}
 The function $\psi_{m}$, is associated with a scalar field~$s_{m}$, as follows 
\begin{equation}\label{mande1}
     \psi_{m}=|\psi^{*}_{m}|\exp\left[\frac{i}{h}s_{m}\right],
 \end{equation}
 in this hydrodynamics the momentum is defined as
 \begin{equation}\label{mande2}
 p^{m}_{\mu}=\partial_{\mu}s_{m}.
  \end{equation}
 We return the Klein-Gordon equation~\ref{mande} and write
  \begin{equation}\label{mande3}
      \partial_{\mu}s_{m}\partial^{\mu}s_{m}-h^{2}\frac{\partial_{\mu}\partial^{\mu}|\psi_{m}| }{|\psi_{m}|}-m^{2}c^{2}=0 .
	\end{equation}
      Since \ref{mande} and \ref{mande3} are a system of Klein Gordon equations in Minkowski space, coupled to conserved current,
      \begin{equation}\label{mande4}
          \frac{\partial}{\partial{q}_{\mu}}\left(|\psi_{m}|^{2}\frac{\partial{S}}{\partial{q}_{\mu}}\right)=m\frac{\partial{J}_{\mu}}{\partial{q}_{\mu}}=0.
      \end{equation}
      
			Starting from \ref{mande1} we get~$s_{m}$ in terms of the massive field~$\psi_{m}$
      \begin{equation}\label{mande4.1}
     s_{m}=\frac{h}{2i}\ln\left[\frac{\psi_{m}}{\psi^{*}_{m}}\right].
      \end{equation}
			
      We write the current
      \begin{equation}\label{mande5}
      J_{\mu}=\left(c\rho,-j_{i}\right)=\frac{ih}{2m}\left(\psi^{*}_{m}\frac{\partial\psi_ {m}}{\partial{q_{\mu}}}-\psi_{m}\frac{\partial\psi^{*}_{m}}{\partial{q}_{\mu}}\right),
      \end{equation}
      which characterizes an incompressible fluid.
			
      We write the 4-momentum in analogy with hydrodynamics
      \begin{equation}\label{mande5.1}
          p^{m}_{\mu}=\left(\frac{E}{c},-p_{i}\right)=-\frac{\partial{s_{\mu}}}{\partial{ q^{\mu}}},
      \end{equation}
      and relate the current to the momentum~\ref{mande3}
      \begin{equation}\label{mande6}
      J_{\mu}=-|\psi_{m}|^{2}\frac{p_{\mu}}{m},
      \end{equation}
      making clear the analogy with hydrodynamics, where the fluid density takes the place of the probability density 
			      \begin{equation}\label{mande61}
\rho=\frac{|\psi|}{mc^2}\frac{\partial{s_{m}}}{ \partial{t}}.
      \end{equation}
			 Considering the equations \ref{mande1} and \ref{mande5.1} we write below
      \begin{equation}\label{mande7}
          \frac{\partial{s}_{m}}{\partial{q}^{\mu}}\frac{\partial{s}_{\mu}}{\partial{q}_{\mu} }=p^{m}_{\mu}p^{\mu}_{m}=\left(\frac{E^2}{c^2}-p^{2}_{m}\right )=m^{2}c^{2}\left(1-\frac{V_{qu}}{mc^2}\right),
      \end{equation}
      where we have that the quantum potential is linked to the massive field $\psi_{m }$ being given by
      \begin{equation}\label{vcu}
          V_{qu}=-\frac{h}{m}\frac{\partial_{\mu}\partial^{\mu}|\psi_{m}|}{\psi_{m}}.
      \end{equation}
       
			The scatter relation expressed in \ref{mande7} is very similar to the scatter relation found in the reference \cite{r3} . However, the scalar potential $\phi$ defined by~\cite{r3}, has a kinematic interpretation associated with the energy-time uncertainty principle, being intimately associated with a clock hypothesis and a deformed metric, while the quantum potential $ V_{qu}$\ref{vcu}, has an interpretation associated with matter waves. Anyway, the equation~\ref{vcu} represents a De-Broglie Bohm type potential. Other authors~\cite{novelo} and~\cite{svt,svt2} interpret the potential~\ref{vcu} in different ways, but in both cases, the interpretations are associated with properties of a conformal geometry, which would act as an effective geometry capable of introducing quantum effects.
 
The cosmological constant is written by~\cite{mand} in terms of the matter fields $\psi_{m}$ and the scalar field $V_{qu}$,
 \begin{equation}\label{vcu2}
     \Lambda=-\frac{mc^2}{\gamma}|\psi_{m}|^{2}\left(1-\sqrt{1-\frac{V_{qu}}{mc^2}} \right),
 \end{equation}
this is yet another point of similarity between the works of~\cite{N}and~\cite{chia}, since the scalar field equation in~\cite{r3} has a functional form very similar to the form of~\ref{vcu2}, and we have already demonstrated in~\cite{r3} that the field $\phi$ associated with the Symmetric Special Relativity (SSR) kinematics has inflationary properties. It is notable that two researchers independently arrived at similar results for the same problem. In the work of~\cite{mand}, the large $\gamma$ has a relation with the line element 
\begin{equation}
\gamma=\frac{c}{\sqrt{g_{\mu\nu}v^{\mu }v^{\nu}}}, 
 \end{equation}
where $v^{\mu}, v^{\nu}$, are canonical 4-speeds.

\subsection{Schutz formalism}
The Schutz formalism is a generalization of the Clebsch formalism~\cite{ufes} where velocity could be written in terms of three potentials $\vec{v}=\vec{\nabla}\phi+\alpha\vec{\nabla} \beta$, so we deal with three scalar potentials $\phi, \alpha, \beta $. The reason for defining this type of vector is that the curl of a gradient is null, but a linear combination of gradients of scalar fields is not null $\vec{\nabla}\times\vec{v}=\vec{\nabla}\alpha\times\vec{\nabla}\beta$, thus eliminating an apparent contradiction in the quantization of and in the definition of vorticity, which we will see in the following sections.

In the Schutz formalism the 4-speed can be written in terms of six scalar fields, thus consisting of a generalization of the Clebsch formalism,
\begin{equation}\label{4v}
 u_{\mu}=k^{-1}\left(\partial_{\mu}\phi+\alpha\partial_{\nu}\beta+\theta\partial_{\mu}s\right) .
\end{equation}

Scalar potentials can be interpreted as thermodynamic potentials, where
\begin{equation}
 k=\frac{\rho+p}{\rho_{0}},
\end{equation}
is the enthalpy. Over the enthalpy the Hawking-Ellis energy conditions \ref{sec}, \ref{wec}, \ref{nec} and the \ref{ndec} condition impose a propagation of signals at infinite speed, this condition being linked to the cosmological constant, so this formalism in a cosmological constant space would imply an infinite enthalpy. We also define the scalar potential $s$ that corresponds to entropy per particle, where the orthogonality relations hold, which links the potentials to each other
\begin{equation}\label{potchi}
    u^{\mu}\partial_{\mu}\phi=-k,
\end{equation} 
where two conservation laws appear
\begin{align}\label{a}
u^{\mu}\partial_{\mu}\alpha&=0 \\\label{b}
u^{\mu}\partial_{\mu}\beta&=0.
\end{align}

We set the temperature
\begin{equation}\label{T}
    u^{\mu}\partial_{\mu}\theta=T,
\end{equation}
then the orthogonality with the entropy gradient, thus defining the isentropic trajectories, in the space of scalar fields
\begin{equation}\label{s}
    u^{\mu}\partial_{\mu}s=0.
\end{equation}

Another link to be established between the potentials derives from the orthogonality property $u^{\mu}u^{\nu}g_{\mu\nu}=-1$,
\begin{equation}
    m^{2}=-g^{\mu\nu}\left(\partial_{\mu}\phi+\alpha\partial_{\mu}\beta+\theta\partial_{\mu}s\right)\left (\partial_{\nu}\phi+\alpha\partial_{\nu}\beta+\theta\partial_{\nu}s\right),
\end{equation}
implying, therefore, that the enthalpy $k$ is the normalizing element. A second order equation, can also be defined considering \ref{4v}, we have
\begin{equation}
    \rho_{0}\nabla^{\mu}u_{\mu}=0.
\end{equation}

The set of equations \ref{potchi}, \ref{a}, \ref{b}, \ref{T} and \ref{s} implies the termination of the following action:
\begin{equation}
    I=\int\sqrt{-g}pd^{4}x,
\end{equation}
here we have $p=p(k,s)$
from which we deduce the first law of thermodynamics as a conserved current.
\begin{equation}
    \delta{P}=\rho_{0}\delta{m}-\rho_{0}T\delta{s}
\end{equation}

\section{The Momentum Energy Tensor}
A Momentum Energy tensor for an ideal fluid in the Schutz formalism.
\begin{equation}
    T_{\mu\nu}=\rho{U}_{\mu}U_{\nu}+P_{\mu\nu}
\end{equation}
here $P_{\mu\nu}=g_{\mu\nu}+U_{\mu}U_{\nu}$
where we will consider for $g_{\mu\nu}$ the acoustic metric \ref{metsom}
\begin{equation}
\mathcal{G}^{\mu\nu}=\frac{\rho}{c}\left(\begin{array}{cc}
c^{2}-v^{2}& -\vec{v}\\
-\vec{v} & \mathcal{I}
\end{array}\right)
\end{equation}

Considering an incompressible fluid $v_{\mu}=\nabla_{\mu}\phi$ where $\phi$ is a scalar field, we then have
\begin{equation}\label{fluid}
    T_{\mu\nu}=(\rho+p)\nabla_{\mu}\phi\nabla_{\nu}\phi-p\mathcal{G}_{\mu\nu},
\end{equation}
so we have a momentum energy tensor for an incompressible fluid. Starting from the Einstein-Hilbert action \ref{eh} ${\int}d^{4}x\sqrt{-g}\left(R+16\pi{p}\right)$ and following \cite{ufes}, we can consider the construction of the Einstein field equation for fluids governed by the tensor energy momentum \ref{fluid}.
\begin{equation}\label{rie1}
\delta\left(\sqrt{-g}R\right)=\left(R_{\sigma\mu}-\frac{1}{2}g_{\sigma\mu}R\right)\sqrt{- g}\delta{g}^{\sigma\nu}
\end{equation}
and
\begin{equation}\label{rie2}
  \frac{\delta{\sqrt{-g}p}}{\delta{g}_{\alpha\beta}}=\frac{1}{2}pg_{\alpha\beta}\sqrt{-g }+\frac{\delta{p}}{\delta\mu}\frac{\delta\mu}{\delta{g}^{\alpha\beta}}\sqrt{-g} .
\end{equation}

We know from \ref{potchi} that 
\begin{align}
\frac{\partial{k}}{\partial{g}^{\sigma\mu}}&=-\frac{k}{2}U_{\sigma}U_{\ mu}\nonumber\\ 
\frac{\partial{p}}{\partial\mu}&=\frac{\rho+p}{k}.\nonumber
\end{align}
Therefore
\begin{equation}
\frac{\delta{\sqrt{-g}p}}{\delta{g}^{\sigma\mu}}=\frac{\sqrt{-g}}{2}\left[pg_{\sigma \nu}+\left(\rho+p\right){k}U_{\sigma}U_{\nu}\right].
\end{equation}
We put together the equations \ref{rie1} and \ref{rie2}, and then arrive at the Einstein field equation for the energy-momentum tensor
\begin{equation}
 R_{\sigma\nu}-\frac{R}{2}g_{\sigma\nu}=\frac{\sqrt{-g}}{2}\left[pg_{\sigma\nu}+\left (\rho+p\right){k}U_{\sigma}U_{\nu}\right],
\end{equation}
hence a field equation for the scalar fields associated with thermodynamic potentials. We therefore find a link between general relativity and thermodynamics.

\section{Superfluids a possible model for antigravity}
Many authors~\cite{volo,vis,zlo,gbecmeu}, have been looking for a model to express what would be antigravity, the cosmological constant~\cite{volo2} and cosmic inflation~\cite{r3} are two theoretical hypotheses that find a phenomenon similar to superfluidity. The properties of superfluids such as energy propagation towards the capillary, the system boundary, could serve as a toy model for a possible antigravity. Models analogous to gravity, especially with superfluids, have received great attention from the scientific community, phenomena of superfluids on the astrophysical scale as in neutron stars~\cite{neutron} and models of ultracompact objects~\cite{gbecmeu2}, also found in superfluidity a phenomenology to be compared.

Let's consider the two-fluid Khalatnikov model~\cite{Huang}, where we have two fluids with different densities $\rho_{s}, \rho_{n}$ and different currents $\vec{j}_{s}$ and $\vec{j}_{n}$ which are orthogonal to each other and both respect the continuity equation \ref{continum}. Then we define the currents
\begin{equation}\label{correescar}
\vec{j}_{s}=\rho_{s}\vec{v}_{s},
\end{equation}
and the perpendicular current
\begin{equation}\label{correperp},
\vec{j}_{n}=\rho_{n}\vec{v}_{n}
\end{equation}
where \ref{correperp} is perpendicular to the flowline~\cite{Huang} and $\vec{j}_{s}$ corresponds to a ring around the flowline.

\subsection{Landau's criteria of superfluidity}
Following \cite{volo,Huang} we will establish the Landau criterion for the existence of a superfluid.
Landau's theory of superfluids is invariant by Galilean transformations between reference frames. Therefore, considering two reference frames~$S$ and $S'$ we have that the momentum is transformed as follows
\begin{equation}\label{chicken}
\vec{P'}=\vec{P}+M\vec{v}.
\end{equation}
In this case, we have $M$ as the total mass of the fluid. We can now write the transformation for the kinetic energy
\begin{equation}\label{tener}
E'=E+\vec{P}\cdot\vec{v}+\frac{Mv^{2}}{2}.
\end{equation}

Therefore we have a fluid motion with constant velocity $\vec{v}$ with respect to the capillary wall. In case any momentum excitation appears, we need to describe it in relation to the referential~$S$, which is in the fluid, therefore a moving referential. We have that the total energy of the fluid is given by
\begin{equation}\label{viagra}
E+\epsilon(p),
\end{equation}
so we have that $E$ is the energy of the ground state and $\epsilon(p)$ is the energy of the excited state, the appearance of $\epsilon(p)$, configures the destruction of the superfluid. In the reference frame~$S$ the capillary wall moves with velocity $-\vec{v}$. The appearance of the excited state modifies \ref{tener} as follows
\begin{equation}\label{tener1}
E'=E_{0}+\epsilon(p)+\vec{p}\cdot{v}+\frac{Mv^{2}}{2}.
\end{equation}
In \ref{viagra} excitations appear under favorable energy conditions, namely
\begin{equation}\label{condiener}
\epsilon(p)+\vec{p}\cdot\vec{v}<0.
\end{equation}

The \ref{condiener} condition implies energy dissipation to the capillary wall, then to the system boundary, thus destroying the superfluid. This condition therefore occurs when we establish a minimum velocity $v_{c}$ for the appearance of excitations,
\begin{equation}\label{vland}
v_{c}=min\frac{\epsilon(\vec{p})}{p}.
\end{equation}

The equation \ref{vland} defines the critical Landau velocity $v_{c}$, the superfluid flow velocity, associated with the first excited degree that we call the roton \cite{Huang}. When dealing with relativistic fluids, the dispersion relation is $\epsilon(p)=\sqrt{p^{2}+M^{2}}$, in this case 
\begin{equation}
\lim_{p\rightarrow0}\frac{d\epsilon(p)}{dp}=0,
\end{equation}
 for minimum values of $p$, then we get that \ref{vland} is null, however, when dealing with linear scatter relations $\epsilon(p)\propto {p }$, we have that 
\begin{equation}
\lim_{p\rightarrow0}\frac{d\epsilon(p)}{dp}=V,
\end{equation}
 a constant value of $V$, so we have that the excited states have a minimum velocity. As a good approximation, we can handle small values of momentum in the scattering relation $\epsilon(p)\propto{p^{2}}$, reducing this scattering relation to a linear relation $\epsilon(p)\propto{ p}$, implying the appearance of a minimum velocity for the excitations. An important classification for the excitations by \ref{viagra} is as follows:
phonos have a linear dispersion relation $\epsilon(p)\propto{p}$, whereas the roton is the first minimum, being characterized by the dispersion relation $\epsilon(p)=\Delta+\frac{(p-p_{ 0})^{2}}{2m}$, in turn, the maxon is the opposite of the roton, being a maximum $\epsilon(p)=\Delta'-\frac{(p-p_{0}) ^{2}}{2m}$. It is possible to interpret the superfluid excitations as another fluid, orthogonal to the first one, so we can use the formulation of~\cite{Huang,volo}, which makes this work much clearer.
In this case, we then write the total density of both fluids as $\rho=\rho_{s}+\rho_{n}$ and the current as $\vec{j}=\vec{j}_{s} +\vec{j}_{n}$, so for both fluids the number of quasiparticles per volume is $n=\sqrt{\rho}$, $\rho$ is the magnitude of the order factor $\psi= \rho e^{\left[i\theta\right]}$.

\subsection{Building collective variables for the components $\vec{j}_{s}$ and $\vec{j}_{n}$}
Let us now study our system in terms of the collective variables $n(\textbf{p})$ which is the density of quasi-particles and $\vec{v}$ the velocity of the fluid~\cite{volo}, we define, according to Volovik \ cite{volo} , the distribution of quasiparticles in thermal equilibrium at a temperature $T$
\begin{equation}\label{qps}
    f_{\tau}(p,v)=\left(\exp\left[\frac{\tilde{E}-\vec{p}\cdot\left(\vec{v}_{n}-\vec {v}_{s}\right)}{T}\right]\pm1\right)^{-1},
\end{equation}
where $+$ represents fermion and $-$ represents Boson.
The scatter relation for $\tilde{E}$ is the same as the \ref{tener1} relation. What makes us able to write, starting from~\ref{qps},
\begin{equation}\label{pm}
\vec{w}=\vec{v}_{s}-\vec{v}_{n}.
\end{equation}

We refer to $\vec{w}$ as counterflow current
\begin{equation}
\vec{j}_{q}=n_{nik}\vec{w}.
\end{equation}

We can now calculate the quasi-particle density, which is given according to~\cite{volo},
\begin{equation}\label{bose}
n_{ni\kappa}=-\Sigma_{p}\frac{p_{i}p_{\kappa}}{m}\frac{\partial{f}_{\tau}}{\partial{E}},
\end{equation}
where $n_{ik}=n\delta_{ik}-n_{nik}$.

\subsection{The quantization of a superfluid}
A superfluid respects the Gross-Pitaevski equation~\cite{Huang}
\begin{equation}\label{gp}
\frac{-\hbar^{2}}{2m}\nabla^{2}\psi+\left(g\psi^{*}\psi-\mu_{0}\right)\psi=i\hbar\frac{\partial\psi}{\partial{t}},
\end{equation}
where $g=4\pi\frac{a\hbar}{m}$.
Although we have discussed in the previous sections an elaboration where a fluid would respect a Klein Gordon equation \ref{mande} with an exponential type solution \ref{mande1}. We had not commented on the possibility of this object representing a superfluid \cite{mand}, we will leave a more in-depth debate on this topic for another opportunity. In this case, we are associating the Schutz formalism with a superfluid  since the two formalisms, Schutz and Mandelung, are the most used to treat quantum potentials. The Mandelung formalism has a dispersion relation \ref{mande7}. Schutz's formalism, on the other hand, allows an immediate connection with thermodynamics, which can be interesting in the treatment of superfluids. In this sense, the solution of the equation \ref{gp} is an exponential $\psi=\rho{e^{\sigma}}$, so we have, as in the Mandelung and Schutz formalism~\ref{4v}, the velocity being described like the gradient of a scalar field
\begin{equation}\label{gradi}
\vec{v}:=\vec{\nabla}\sigma.
\end{equation}

The quantization of the velocity, however, occur when we do the line integral
\begin{equation}
\int\vec{v}\cdot\textbf{ds}.
\end{equation}
Usually, for a gradient vector this integral would be zero, but here we write
\begin{equation}\label{vortex}
    \int\vec{v}\cdot\textbf{ds}=2n\pi\hbar,
\end{equation}
where $n=0,1,2,3,...$ Implying that velocity is associated with a vortex and
\begin{equation}\label{ampere}
    v=\frac{2\pi\hbar{n}}{mr},
\end{equation}
by~\cite{vis}, $\vec{v}$ obeys the equations \ref{eu1}, \ref{eu2}, being able to be a model for the dark matter \cite{mecitou}. But the issue is really the quantization of velocity, it is usually expected that $
    \int\vec{\nabla}\sigma\cdot\textbf{ds}=0$, the fact that \ref{vortex} is not null have enigmatic consequences, which are still the subject of intense study~\cite{Huang}. One of the open discussions is the relation with the simple connectedness of the space \cite{Huang2,sup}. So the vector~\ref{gradi} cannot be written solely as a gradient. Having thus a component that allows us to write $\vec{\nabla}\times\vec{v}\neq0$, following \cite{Huang} we write 
		\begin{equation}
    \vec{v}=\vec{\nabla}\phi+\vec{v}_{ind},
    \end{equation}
    it offers us
    \begin{equation}
    \vec{\nabla}\times\vec{v}=\vec{\nabla}\times\vec{\nabla}\phi+\vec{\nabla}\times\vec{b}.
    \end{equation}
    
We define the vorticity
\begin{equation}
\vec{k}=\vec{\nabla}\times\vec{v}_{ind}.
\end{equation}
    Considering a vector tangent to the flowline of the vortex $\vec{s}(\xi,t)$, where $\xi$ is a parameter that runs on the flowline itself and $t$o time. In this sense we can write the vorticity in terms of this vector
    \begin{equation}\label{vorticity}
        \vec{k}(\vec{r},t)=k_{0}\int{ds}\delta\left(\vec{r}-\vec{s}(\xi,t)\right),
    \end{equation}
    the enormous similarity with electromagnetism, suggests that the velocity $\vec{b}$, must be written by the Biort-Savat law, being $\vec{b}$ associated with the rotation velocity of the vortex around the line of flow.
    \begin{equation}\label{b2}
    \vec{v}_{int}:=\frac{k}{4\pi}\int{d\xi'}\frac{(\vec{s}(\xi',t)-\vec{s }(\xi,t))}{|\vec{s}(\xi',t)-\vec{s}(\xi,t)|^{3}},
    \end{equation}
    where $\vec{s}_{1}$ is a particular point on the flowline and extends to all points on the flowline 
  \begin{equation}
	d\vec{s}(\xi,t)=\frac{\partial \vec{s}}{\partial\xi}d{\xi}+\frac{\partial\vec{s}}{\partial{t}}dt,    
	\end{equation}
		 where 
		  \begin{align}
	\vec{s}'&=\frac {\partial\vec{s}}{\partial\xi}\nonumber\\
	\dot{\vec{s}}&=\frac{\partial\vec{s}}{\partial{t}}.\nonumber
	\end{align}
	We can relate $\vec{b}$,$\dot{\vec{s}},\vec{s}'$, 
    \begin{equation}
      \dot{\vec{s}}=\vec{v}_{ind}(\vec{s})+\vec{v}+\frac{\alpha}{|\vec{s}'|}\ vec{s}'\times\left(\vec{v}_{ns}-\vec{v}_{ind}\right)-\alpha'\frac{\vec{s}''}{|\ vec{s}'|}\times\left[\frac{\vec{s}'}{|\vec{s}'|}\times\left(\vec{v}_{ns}-\vec{ v}_{ind}\right)\right],
    \end{equation}
  with the module $|\vec{s}''|=\frac{1}{R}$, where $R$ is the radius of the circulation ring, so there is a link between the vector $\vec{s}'' $ is the speed modulus given by \ref{ampere}. Knowing that the vectors $ \vec{s}', \vec{s}'', \vec{s}'\times\vec{s''}$ generate a moving base like Frenet-Serrat~\cite{fs}, then we also obtain a relation of the circulation velocity with this basis.
    
		Returning to \ref{io} and making $\dot{\vec{s}}(\xi,t)\equiv\vec{V}_{L}(\xi,t)$, as well as $\vec{v }_{s}(\vec{s}(\xi,t), t)\equiv\vec{V}_{sl}(\xi,t)$ we write the Magnus force, associated with the friction between the layers of the fluid,
    \begin{equation}\label{mag}
        \frac{f_{m}}{\rho_{s}k_{0}}=\frac{\vec{s}}{|\vec{s}'|}\times\left(\vec{v}_ {L}-\vec{v}_{sl}\right).\end{equation}
        The magnus force \ref{mag}, indicates the importance of Galilean symmetry in this fluid.
				
\subsection{Iordenski force}
After establishing the Galinean symmetry for the excitations of superfluids \ref{chicken}, we will study the appearance of a force between these excited states, which is known as the Iordenski force \cite{Huang,sup},
\begin{equation}\label{io}
\vec{F}_{io}=-D\left[e_{z}\times\left(\vec{u}_{e}-V_{L}\right)\right]+D'\left[ e_{z}\times\left(\vec{u}_{e}-\vec{V}_{L}\right)\right].
\end{equation}

Here we have that $e_{z}$ is defined by the direction of a cyon type system~\cite{sup}, where $\vec{\Phi}=\phi{e_{z}}$, the velocity $V_{L} $ is the velocity of the vortex. Therefore, we consider the particles to be vortices, so they would be better identified as quasiparticles, thus having an internal structure, $u_{e}$ is the propagation velocity of the particle. We note that there is a Galilean symmetry $\vec{u}_{e}-\vec{V}_{L}$, so this force is, in a good sense, a force dependent on a Galilean transformation. The Iordenski force is largely similar to the Lorentz force and the field $\vec{\Phi}$ carries an Aharonov-Bohm \cite{sup} structure at its origin. However, for developments in the next sections, our interest is in the relation between the iordenski force and Galilean symmetry. The Iordeski force brings a novelty, which is a force associated with the conservation of Galilean symmetry, we have a different perspective of the Galilean systems, where the usual Galilean symmetry is valid.
    
		\subsection{Relativistic two fluids Hydrodynamycs -- An application to the Schutz formalism}
    Following \cite{Huang2} we define an ideal fluid based on the superfluids defined above \ref{pm},
\begin{equation} 
\Pi^{jk}=\rho_{s}v^{j}_{s}v^{k}_{s}+\rho_{n}v^{j}_{n} v^{k}_{n}+p\delta_{jk},
\end{equation}
and its conservation
\begin{equation}
    \frac{\partial{j}^{k}}{\partial{t}}+\partial_{j}\Pi^{jk}=0.
\end{equation}

We introduce a gauge transformation
\begin{align} 
\frac{\partial}{\partial{t}}&\rightarrow\nabla-i\phi,\\
\vec{\nabla}&\rightarrow\vec{\nabla}-i\vec{A},
\end{align}
we then write a Lagrangian that respects the nonlinear Schroedinger equation
\begin{equation}
\mathcal{L}=\mathcal{L}_{0}+\rho\phi-\vec{j}\cdot\vec{A}+\frac{\rho}{2}A^{2}.
\end{equation}
In this Lagrangian we have the validity of Galileo's symmetry.
\begin{align}
    \vec{A}&=\alpha\left(\vec{v}_{s}-\vec{v}_{n}\right),\\
\phi&=\vec{v}_{n}\cdot\vec{A}.
\end{align}

Constituting itself as a fundamental symmetry for the very emergence of the excitations of the superfluid \ref{qps}, according to the criterion of Landau \ref{vland}. Such symmetries configure a potential
\begin{equation}
U=-\frac{1}{2}\left(\vec{v}_{n}-\vec{v}_{s}\right)\frac{\partial\rho_{n}}{\partial \rho}-\frac{1}{2\rho}\vec{\nabla}\left[\rho_{n}\left(\vec{v}_{n}-\vec{v}_{s} \right)\right],
\end{equation} which appears in the nonlinear Schroedinger equation, also known as the Gross Pitaevski equation \ref{gp}
\begin{equation}\label{nlse}
i\partial_{t}\psi=-\frac{1}{2}\nabla^{2}\psi+\left(|\psi|^{2}-1+U\right)\psi.
\end{equation}

We will now see a perturbation in Galileo's symmetry, so we start to study superfluids no longer described by the nonlinear Schroedinger equation.

\subsection{NLKG-Mandelung's formalism and superfluids}
    For a nonlinear Klein Gordon equation, we define a complex scalar field as we did in the Mandelung formalism \ref{mande2}
\begin{equation}\label{order}
\phi(\vec{x},t)=F(\vec{x},t)e^{[i\sigma(\vec{x},t)]}.
\end{equation}
The flow in the tensor formalism, for example~\ref{4v}
\begin{equation}\label{347}
    v^{\mu}_{s}=\partial^{\mu}\sigma,
\end{equation}
the associated trivector
\begin{equation}
\vec{v}_{s}=\frac{\vec{\nabla}\sigma}{\partial^{0}\sigma},
\end{equation}
we have mandatory that $\frac{|\vec{v}_{s}|}{c}<1$.

We write the action for a scalar field
\begin{equation}\label{snlkg}
\mathcal{L}=\int{d^{4}x}\sqrt{-g}\left(g^{\mu\nu}\partial_{\mu}\phi^{*}\partial_{\nu }\phi+V\right),
\end{equation}
the conserved current generated by \ref{snlkg} is
\begin{equation}
\left(\square-V'\right)\phi=0
\end{equation}
where 
\begin{align}
V'&=\frac{d\mathcal{L}}{d(\phi^{*}\phi)}\\
\square\phi&\equiv\frac{1}{\sqrt{-g}} \partial_{\mu}\left(\sqrt{-g}g^{\mu\nu}\partial_{\nu}\phi\right).
\end{align}

We then write the hydrodynamic equations for \ref{order}
\begin{align}
\left(\square-V'\right)F+F\nabla^{\mu}\sigma\nabla_{\mu}\sigma&=0,\\
2\nabla^{\mu}F\nabla_{\mu}\sigma+F\nabla^{\mu}\nabla_{\mu}\sigma&=0,
\end{align}
so we define the current associated with~\ref{order}
\begin{equation}
j^{\mu}=\frac{1}{2i}\left(\phi^{*}\partial^{\mu}\phi-\phi\partial^{\mu}\phi^{*}\right)=F^{2}\partial^{\mu}\sigma .
\end{equation}

The similarity of the action \ref{snlkg} with the action of a k-essence model \cite{kesse} is remarkable.
As an example of \ref{4v} we now define
\begin{equation}
w^{\mu}:=\partial^{\mu}\alpha+\xi\partial^{\mu}\beta,
\end{equation}
here $\alpha,\xi,\beta$ are independent variables that respect the Schutz formalism \ref{potchi}, \ref{a}, \ref{b}, \ref{T}, \ref{s} with these variables we can write some Lorentzian invariants and it still solves the problem of the contradiction between the definition of the velocity vector as a gradient \ref{gradi} and the velocity to Ampere's law \ref{ampere}. For this case $\partial_{ [\mu}w_{\nu]}\neq0$,
\begin{align}
I_{1}&=\frac{F^{2}}{2}v^{\mu}v_{\nu},\\
I_{2}&=F^{2}v^{\mu}w_{\mu},\\
I_{3}&=\frac{F^{2}}{2}w^{\mu}w_{\mu}.
\end{align}

We write the extended Lagrangian with respect to \ref{snlkg}
\begin{equation}
\mathcal{L}=\mathcal{L}_{0}+\mathcal{F}(I_{1},I_{2}, I_{3}).
\end{equation}
We explicitly write the Schutz formalism
\begin{equation}
\left[\square-V'-(1+f'_{1})v_{\mu}v^{\mu}-2f'_{2}w_{\mu}v^{\mu}-f '_{3}w_{\mu}w^{\mu}\right]F=0
\end{equation}
with $f'=\frac{\partial\mathcal{F}}{\partial{I}_{n}}$. We still have other conserved currents
\begin{align}
\partial_{\mu}j^{\mu}&=0,\\
\partial_{\mu}s^{\mu}&=0,\\
s^{\mu}\partial_{\mu}\xi&=0,\\
s^{\mu}\partial_{\mu}\beta&=0,
\end{align}
fitting the same thermodynamic interpretation of the equations \ref{potchi}, \ref{a}, \ref{b}, \ref{T} and \ref{s}.

Here we return to the equation~\ref{347}
\begin{equation}\label{3471}
v^{\mu}=(\tilde{\mu},\vec{\nabla}\sigma)=\tilde{\mu}\left(1, \vec{v}_{s}\right)
\end{equation} and a perpendicular fluid
\begin{equation}\label{3472}
v^{\mu}_{n}=\tilde{s}\left(1,\vec{v}_{n}\right)
\end{equation}
we establish normalization
\begin{equation}
\tilde{\mu}=\gamma_{s}\mu,
\end{equation}implying a Lorentz factor
\begin{equation}
\gamma_{s}=\frac{1}{\sqrt{1-\frac{v^{2}_{s}}{c^{2}}}}
\end{equation}
the same for $\tilde{s}$
\begin{equation}
\tilde{s}=\gamma_{n}s
\end{equation}
\begin{equation}
\gamma_{n}=\frac{1}{\sqrt{1-\frac{v^{2}_{n}}{c^{2}}}}.
\end{equation}

The dot product between \ref{3471} and \ref{3472} is
\begin{equation}
v^{\mu}_{s}v^{n}_{\mu}=-c^{2}y=-c^{2}\tilde{\mu}\tilde{s}\left(1 -\frac{\vec{v}_{s}\cdot\vec{v}_{n}}{c^{2}}\right)=-c^{2}\mu{s}\gamma_{ n}\gamma_{s}\left(1-\frac{\vec{v}_{s}\cdot\vec{v}_{n}}{c^{2}}\right),
\end{equation}
the Lorentz transformations for the addition of $\vec{v}_{s}$ and $\vec{v}_{n}$
\begin{equation}
\vec{v}_{ns}=\frac{\vec{v}_{n}-\vec{v}_{s}}{1-\frac{\vec{v}_{s}\cdot \vec{v}_{n}}{c^{2}}}
\end{equation}
so the squared modulus $||\vec{v}_{ns}||=||\vec{v}_{s}-\vec{v}_{n}||^{2}$ is calculated as
\begin{equation}
\frac{v^{2}_{ns}}{c^{2}}=1-\frac{\mu^{2}s^{2}}{y^{4}}.
\end{equation}
It is established here a relation of the product of the modulus of the difference between the perpendicular and circulation velocities $||\vec{v}_{s}-\vec{v}_{n}||^{2}$ with the entropy of the system.

\section{Superfluid Vacuum theory -- a brief review}
In this section, we will study a formalism that combines superfluids with effective metrics, see~,\cite{volo}, we consider the linear phonon spectrum in the momentum
\begin{equation}
E(p,n)\rightarrow{c}(n)|\vec{p}|
\end{equation}
where $c(n)$ is given in \ref{ene}. Phonons represent the collective quanta of the excited modes of a superfluid, whose velocity is given by
\begin{equation}
c^{2}(n)=\frac{n}{m}\left(\frac{d^{2}\epsilon}{dn^{2}}\right).
\end{equation}
Following the concept of acoustic metric~\ref{metsom}, we can associate the geometric structure with hydrodynamic quantities,
\begin{align}
g^{00}&=-\frac{1}{mnc},\\
g^{0i}&=-\frac{v^{i}_{s}}{mn^{2}},\\
g^{ij}&=\frac{c^{2}\delta^{ij}-v^{i}_{s}v^{j}_{s}}{mnc},\\
g_{00}&=-\frac{mn}{c}\left(c^{2}-v^{2}_{s}\right),\\
g_{0i}&=\frac{mn}{c}v_{si},\\
\sqrt{-g}&=\frac{m^{2}n^{2}}{c^{2}}.
\end{align}

Then the components of the metric $g_{\mu\nu}$ are associated with the density of states $n$, the mass $m$, the modulus of the fluid velocity $v$, the modulus of the propagation of signals $c$ and to the velocity vector of the flow line $v_{si}$.
The ``gravitons'' of this acoustic geometry respect the geodesic
\begin{equation}
g^{\mu\nu}p_{\mu}p_{\nu}=0,
\end{equation}
and the dispersion relation
\begin{equation}
    \left(\tilde{E}-\vec{p}\cdot\vec{v}_{s}\right)^{2}=(cp)^{2}.
\end{equation}

\subsection{Mandelung formalism applied to dark matter}
We use \ref{mande1} to describe quantum vacuum fluctuations, interpreting \ref{mande1} as a wavefunction $\Psi(\textbf{r},t)$, in three-dimensional Euclidean space. Building then the normalization condition~\cite{zlo}
\begin{equation}
<\psi|\psi>=\int_{\nu}\rho{d\nu}=\mathcal{M},
\end{equation}
where $\mathcal{M}$ is the total mass and $\nu$ is the volume of a fluid, so we write the mass density of the fluid $\rho=|\psi|^{2}$. The function of \ref{mande1} is governed by the Schroedinger equation
\begin{equation}
\left[-i\hbar\partial_{t}-\frac{\hbar^{2}}{2m}\nabla^{2}+V_{ext}(\vec{r},t)+F\left (|\psi|^{2}\right)\right]\psi=0,
\end{equation}
where $m$ is the mass of the constituent particle, $V_{ext}(\vec{r},t)$ is an external potential and $F(\rho)$ is a properly chosen function. The wave function can be formally derived as a minimizing condition of a functional action that follows the Lagrangian
\begin{equation}\label{zlo}
\mathcal{L}=\frac{i\hbar}{2}\left(\psi\partial_{t}\psi^{*}-\psi^{*}\partial_{t}\psi\right)+ \frac{\hbar^{2}}{2m}|\nabla\psi|^{2}+V_{ext}(\vec{r},t)|\psi|^{2}+V(|\ psi|^{2}).
\end{equation}

The excitations governed by~\ref{zlo}, can be understood as a fluid analogous to gravity, thus having an induced geometry as defined in \ref{metsom}, which we rewrite below, considering the velocity as given in \ref{mande2}
\begin{equation}\label{I found2}
\mathcal{G}^{\mu\nu}=\frac{\rho}{c}\left(\begin{array}{cc}
-[c^{2}_{s}-\eta^{2}(\nabla{S})^{2}] & -\eta\vec{\nabla}s\\
-\eta\vec{\nabla}s & \mathcal{I}
\end{array}\right),
\end{equation}
the factor being $\eta=\frac{\hbar}{m}$ and $s=s(\vec{r},t)=-i\ln\left(\frac{\psi(\vec{r} ,t)}{|\psi(\vec{r},t)|}\right)$, we still need to establish that $c_{s}$ is the speed of sound. In Mandelung's representation $ \psi=\sqrt{\rho}e^{is}$ is a unitary matrix in three dimensions. We also established that $c_{s}>\eta|\nabla{s}|$, we are talking here, therefore, about small oscillations of small amplitude of a condensate, moving along the induced geodesic, as in \ref {geo}. In this example, authors like \cite{svt} consider that waves propagate with the speed of sound in a material medium, therefore $c_{s}=\sqrt{\frac{dp}{d\rho}}$, thus establishing , that space-time behaves as a material medium under certain conditions, we refer here to what \cite{svt} calls fluid/gravity, in this formulation it is not possible to remove the backgroud, and the metric itself corresponds to matter distribution, this fact produces a slightly different interpretation of the Einstein field equation \ref{campo} deduced from the Einstein-Hilbert action gains a new interpretation here. For both the left side corresponds to matter and the right side to a geometry induced by matter
\begin{equation}\label{efe}
T^{\mu\nu}_{induces}{\equiv}R^{\mu\nu}-\frac{R}{2}g^{\mu\nu}.
\end{equation}

So the difference between \ref{campo} and \ref{efe} is that \ref{campo} is understood as a differential equation for unknown terms of the metric. On the other hand \ref{efe} corresponds to a coupling of the test particles, where the condition of small oscillations is not violated. Another difference is that \ref{efe} is seen by R-Observer. In this formalism called fluid-gravity we define two types of observers: the R-observer that is able to observe small measurements, small moments, small deviations and an F-observer that is in the Euclidean space \cite{svt2}. This approximation can be applied to dark matter~\cite{mecitou}. This entire formulation respects Eulerian relations, which are crystallized in the continuity equation
\begin{equation}\label{eu1}
\frac{dn}{dt}+\vec{\nabla}\cdot\vec{v}=0,
\end{equation} and in Euller's equation
\begin{equation}\label{eu2}
m\frac{d\vec{v}}{dt}=m\left[\partial_{t}\vec{v}+(\vec{v}\cdot\vec{\nabla})\vec{v} \right]=-\vec{\nabla}\left[U(\vec{r})+U_{rot}(\vec{r})+u_{0}n-\frac{\hbar^{2} }{2m}\frac{\Delta\sqrt{n}}{\sqrt{n}}\right],
\end{equation}
and the Thomas-Fermi approximation
\begin{equation}\label{tf}
\nabla\left[U(\vec{r})+U_{rot}(\vec{r})+u_{0}n\right]=0.
\end{equation}

In \ref{tf} we have that $n=\psi(\vec{r},t)$
\begin{equation}\label{wheel}
U_{rot}(\vec{r})=-w^{2}r^{2},
\end{equation}
where $w$ is the rotation speed of the dark matter Galactic Halo \cite{mecitou}. The equilibrium of the superfluid associated with dark matter is given by
\begin{equation}
    \Delta\rho(\vec{r})+k^{2}\left[\rho(\vec{r})+\rho_{b}(\vec{r})+\rho_{disc}(\vec{r})-\frac{w^{2}}{2G\pi}\right]=0,
\end{equation}
here $k^{2}=\frac{Gm^{3}}{a\hbar^{3}}=\frac{4{\pi}Gm^{2}}{u_{0}}$, $ \rho=mn(\vec{r},t)$, $a$ is the scattering length in the equation \ref{tf}, we have $U(\vec{r})$ being the gravitational potential, which is usually worked out as having three parts, associated with the gravitational potential of baryonic matter, dark matter and the system disorder~\cite{mecitou}.

\section{Conclusions}
  The study of the momentum-energy tensor is a fundamental topic in relativity, the review of classical conditions of
energy is a necessity given its relevance to several lines of research in astrophysics and cosmology. The emergence of dark and exotic relativistic fluids, brings the need to review their energy conditions especially in terms of causality.
The anisotropy of the momentum-energy tensor is a topic that still raises many doubts among students and the
cosmological constant is increasingly being associated with ultra-dense objects. We review these themes by calculating the conservation
of the momentum-energy tensor, we seek to contribute to the discussion about energy conditions,
both in the isotropic and anisotropic case, identifying exactly which energy condition is associated with vacuum energy.
%
   %
%
%
%
   %
   %
   %
%
%

We built the superfluid structure via Landau formalism, introduced Galilean and Lorentzian symmetries and some of its aspects applied to superfluids were discussed.

In the future, we intend to extend this study to a review of the SVT theory, also to study relativistic fluids associated with nuclear astrophysics such as Bjorken and Gupsen fluids and review properties of relativistic fluids with viscosity, in addition to investigating further the causal structure, especially in dark and tachionic fluids. Exploring fluid-gravity correspondence is also one of our goals.

%

%
%
\bibliographystyle{apsrev4-1}
\bibliography{referencias}

\end{document}